\documentclass[fleqn,10pt]{wlscirep}

\usepackage[utf8]{inputenc}
\usepackage[T1]{fontenc}

\usepackage{graphicx}
\usepackage{multirow}
\usepackage{color}
\usepackage{bm}
\usepackage{amsfonts,amssymb,amsmath}
\usepackage{graphicx,dcolumn,bm,color,braket,slashed}
\usepackage{times}
\usepackage{comment}
\usepackage{cleveref}
\usepackage{pdfpages}

\def\t#1{\textrm{#1}}

\newcommand\abs[1]{\left|#1\right|}

\title{Emergence of Topological Superconductivity in Doped Topological Dirac Semimetals under Symmetry-Lowering Lattice Distortions}

\author[1,2,3,4, +]{Sangmo Cheon}
\author[1,2,5,+]{Ki Hoon Lee}
\author[1,2,6,7,*]{Suk Bum Chung}
\author[1,2,8,*]{Bohm-Jung Yang}

\affil[1]{Center for Correlated Electron Systems, Institute for Basic Science (IBS), Seoul 08826, Korea}
\affil[2]{Department of Physics and Astronomy, Seoul National University, Seoul 08826, Korea}
\affil[3]{Department of Physics and Research Institute for Natural Science, Hanyang University, Seoul 04763, Korea}
\affil[4]{Institute for High Pressure, Hanyang University, Seoul 04763, Korea}
\affil[5]{Department of Physics, Incheon National University, Incheon, 22012, Korea}
\affil[6]{Department of Physics, University of Seoul, Seoul, 02504, Korea}
\affil[7]{Natural Science Research Institute, University of Seoul, Seoul, 02504, Korea}
\affil[8]{Center for Theoretical Physics (CTP), Seoul National University, Seoul 08826, Korea}
\affil[*]{sbchung0@uos.ac.kr; bjyang@snu.ac.kr}
\affil[+]{these authors contributed equally to this work}

\begin{abstract}
Recently, unconventional superconductivity having a zero-bias conductance peak is reported in doped topological Dirac semimetal (DSM) with lattice distortion.
Motivated by the experiments, we theoretically study the possible symmetry-lowering lattice distortions and their effects on the emergence of unconventional superconductivity in doped topological DSM.
We find four types of symmetry-lowering lattice distortions that reproduce the crystal symmetries relevant to experiments from the group-theoretical analysis.
Considering inter-orbital and intra-orbital electron density-density interactions, we calculate superconducting phase diagrams. We find that the lattice distortions can induce unconventional superconductivity hosting gapless surface Andreev bound states (SABS). 
Depending on the lattice distortions and superconducting pairing interactions, the unconventional inversion-odd-parity superconductivity can be either topological nodal superconductivity hosting a flat SABS or topological crystalline superconductivity hosting a gapless SABS. 
Remarkably, the lattice distortions increase the superconducting critical temperature, which is consistent with the experiments.
Our work opens a pathway to explore and control pressure-induced topological superconductivity in doped topological semimetals.
\end{abstract}
\begin{document}

\flushbottom
\maketitle
\thispagestyle{empty}
\section*{Introduction}
Topological insulator, Dirac semimetal (DSM), Weyl semimetal, and topological superconductor are newly established quantum states of matter which are expected to have applications for dissipationless devices and quantum information technologies\cite{hasan_colloquium:_2010, qi_topological_2011,book_topological_2013,chiu2016classification,ando2015topological,armitage_weyl_2018,sato_topological_2017}.
Among them, topological Weyl and Dirac semimetals are characterized by relativistic quasi-particles and gapless nodes in bulk spectra\cite{wan2011topological,burkov2011topological,young2012dirac, book_topological_2013,armitage_weyl_2018,hasan2021weyl}.
Because of their anomalous electromagnetic responses and topologically-protected surface Fermi arcs on the boundaries, such topological semimetals have been attracted much attention
\cite{nielsen1983adler,murakami2008universal,wan2011topological,yang2011quantum,son2013chiral,kim2013dirac,hosur2013recent,vafek2014dirac,okugawa2014dispersion,liang2015ultrahigh,yang2015weyl,lv2015experimental,burkov2016z,yan_topological_2017,wang_quantum_2017,armitage_weyl_2018,nagaosa_transport_2020}.
Moreover, due to the unique properties of Dirac and Weyl semimetals, extensive theoretical and experimental studies of their superconducting instabilities have been conducted to observe possible topological superconductivity\cite{ando2015topological,sato_topological_2017}.

Recently, the lattice-distortion induced superconductivity in DSMs of Cd$_3$As$_2$\cite{aggarwal_unconventional_2016,he_pressure-induced_2016,wang2016observation} and Au$_2$Pb\cite{schoop_dirac_2015,chen_temperature-pressure_2016,xing_superconductivity_2016,yu2017fully,wu2019ground} is reported.
For Cd$_3$As$_2$, it does not show any superconductivity at the ambient pressure until 1.8 K\cite{aggarwal_unconventional_2016,he_pressure-induced_2016,wang2016observation}.
The structural phase transition occurs near 2.6 GPa from 
a tetragonal lattice with $D_{4h}$ point group symmetry ($I4_1/acd$) to a monoclinic lattice with $C_{2h}$ point group symmetry ($P2_1/c$).
Then, superconductivity emerges at $T_c \approx 1.8$ K under pressure higher than 8.5 GPa.
When the pressure increases further, $T_c$ keeps increasing from 1.8 K to 4.0 K in the hydrostatic pressure experiment\cite{he_pressure-induced_2016}.
Similarly, Au$_2$Pb shows superconductivity at $T_c \approx 1.2$ K after a structural phase transition from the cubic with $O_h$ symmetry ($Fd3m$) to the orthorhombic lattice with $D_{2h}$ symmetry ($Pbnc$)\cite{schoop_dirac_2015,xing_superconductivity_2016,wu2019ground}. 
$T_c$ increases up to 4~K at 5 GPa, then decreases with further compression\cite{wu2019ground}. For both materials, the point-contact measurements reported that measured $T_c$ using a hard contact tip is much higher than the measured $T_c$ using a soft tip\cite{aggarwal_unconventional_2016, wang2016observation,xing_superconductivity_2016}.
The point-contact measurements for Cd$_3$As$_2$ showed the zero-bias conductance peak (ZBCP) and double conductance peaks symmetric around zero bias, which was interpreted as a signal of a topological Majorana surface state\cite{aggarwal_unconventional_2016,wang2016observation}.
Moreover, the transport data under magnetic fields reported anomalous behaviors that the conventional BCS theory cannot explain \cite{aggarwal_unconventional_2016,wang2016observation,xing_superconductivity_2016}.
At ambient pressure, the proximity-induced superconductivity in Cd$_3$As$_2$ is also reported \cite{yu_zero-bias_2020}.

In parallel to the experimental exploration of the superconductivity in doped DSM, several theoretical studies were conducted\cite{kobayashi2015topological,hashimoto2016superconductivity}.
In the absence of lattice distortion, the possible superconducting states in doped DSM are suggested as either fully-gapped superconductor (FGSC) or topological nodal superconductor (TNSC) hosting a flat surface Andreev bound state (SABS) on the boundary\cite{hashimoto2016superconductivity}.
In experiments, however, superconductivity was observed only in the presence of lattice distortion.
Considering a lattice distortion (in our work, $n_1$ type lattice distortion), the topological crystalline superconductor (TCSC) hosting surface Majorana states was proposed\cite{kobayashi2015topological}.
However, because such lattice distortion results in the orthorhombic lattice,
it cannot be applied to the observed superconductivity in the monoclinic crystal structure of Cd$_3$As$_2$\cite{he_pressure-induced_2016}.
It is, therefore, necessary to study the effect of symmetry-lowering lattice distortions on the emergence of unconventional superconductivity in doped DSM.

In this work, we systematically study possible symmetry-lowering lattice distortions and their effects on the emergence of unconventional superconductivity in doped topological DSM.
As a representative model, we consider a topological DSM described by the four-band Hamiltonian having $D_{4h}$ point group symmetry in the absence of lattice distortions. 
While keeping time-reversal symmetry (TRS) and inversion symmetry (IS), we find four types of symmetry-lowering lattice distortions from the group-theoretical analysis, which are denoted as $n_i$ type lattice distortions ($i=1,\ldots,4$).
Two of them ($n_1$ and $n_2$ type) reduce $D_{4h}$ of the tetragonal lattice to $D_{2h}$ orthorhombic lattice, while the others ($n_3$ and $n_4$ type) transform the tetragonal lattice to $C_{2h}$ of the monoclinic lattice.
They explain the structural phase transition in Cd$_3$As$_2$ and Au$_2$Pb under pressure. The symmetry-lowering lattice distortions are summarized in Table~\ref{table:classification_distortion_leading}.

To understand the emergence of superconductivity under lattice distortions, we adopt the Bogoliubov-de Gennes (BdG) formalism and linearized gap equation, and we assume intra-orbital ($U$) and inter-orbital ($V$) electron density-density interactions which induce superconducting instabilities.
From the Fermi-Dirac statistics, six possible momentum-independent superconducting pairing potentials are found\cite{hashimoto2016superconductivity}.
Under lattice distortions, six pairings potentials are classified according to the irreducible representation of the remaining point symmetry group.
Using these pairing potentials, possible superconducting gap structures and superconducting critical temperatures ($T_c$) are calculated.
By comparing critical temperatures, we obtain the superconducting phase diagram, and the dominant superconducting phases are discovered, such as 
fully-gapped superconductor (FGSC), topological nodal superconductor (TNSC), and topological crystalline superconductor (TCSC) depending on the lattice distortions and the ratio of $U/V$.
Among them, FGSC is conventional superconductor, while TNSC and TCSC are unconventional.

Interestingly, the unconventional superconductors of TNSC and TCSC emerge when inter-orbital interaction $V$ and the strength of lattice distortion are large enough while FGSC emerges in the opposite limit.
Therefore, the lattice distortion and inter-orbital interaction act as physical parameters that control the phase transition between conventional and unconventional superconductivity of a topological DSM.
We find that both $V$ and lattice distortions enhance the unconventional superconducting pairings via a unique spin-orbit locking.
Moreover, $T_c$ increases under the lattice distortions due to the enhancement of DOS at the Fermi surface, which is consistent with the experimentally measured $T_c$ enhancement under pressure.
The unconventional superconductors host gapless SABS in mirror plane even under the lattice distortions:
Under the $n_1$ or $n_2$ type lattice distortion, the superconductivity in the orthorhombic lattice with $D_{2h}$ point group symmetry hosts a gapless SABS protected by the mirror Chern number.
Under the $n_3$ or $n_4$ type lattice distortion,
the superconductivity in the monoclinic lattice with $C_{2h}$ point group symmetry hosts a gapless SABS protected by the unbroken mirror symmetry and a flat SABS protected by the mirror chiral winding number in specific conditions.
Because there exist gapless Majorana surface states under the lattice distortions, we suggest that these states can be observed in scanning tunneling microscope (STM) or point contact Andreev reflection spectroscopy experiments.

Consequently, our theoretical work is consistent with
the discovered structural phase transition and the enhancement of superconductivity in Cd$_3$As$_2$ and Au$_2$Pb under lattice distortions.
Moreover, we suggest that the emergence of conventional and unconventional superconductivity in doped topological DSM can be controlled by the pressure and strength of the superconducting pairing interaction.
Therefore, our woks opens a pathway to explore and control the topological superconductors in doped topological semimetals, which may have future applications in dissipationless and quantum information devices.

\section*{Results}
\subsection*{Undistorted Dirac semimetal}

Dirac semimetal (DSM) has the low energy excitations near the Fermi-level described by a massless Dirac equation. 
Because all bands are doubly degenerate due to the TRS and IS, a DSM is minimally described by a four-band Hamiltonian\cite{murakami2007phase,yang2014classification,gao2016classification,armitage_weyl_2018,young2012dirac}.
However, TRS and IS are not enough to protect a fourfold degeneracy, so the symmetry-protected DSM is suggested,
where the Dirac points are protected by TRS, IS and crystalline symmetries\cite{murakami2007phase,young2012dirac,yang2014classification,gao2016classification,armitage_weyl_2018}.
DSMs are reported in many materials such as $\beta$-cristobalite BiO$_2$\cite{young2012dirac}, distorted spinels \cite{steinberg2014bulk}, Na$_3$Bi\cite{liu2014discovery,xu2015observation}, Cd$_3$As$_2$\cite{wang2012dirac,wang2013three,neupane2014observation,jeon2014landau,liu2014discovery,liu2014stable,borisenko2014experimental}, Au$_2$Pb \cite{schoop2015dirac,chen2016temperature}, and ZrTe$_5$\cite{chen2015optical,li2016chiral}.
Among them, superconductivity is reported in  Cd$_3$As$_2$\cite{aggarwal_unconventional_2016,he_pressure-induced_2016,wang2016observation} and Au$_2$Pb\cite{schoop_dirac_2015,chen_temperature-pressure_2016,xing_superconductivity_2016,yu2017fully,wu2019ground}.
Both materials have Dirac points protected by TRS, IS, and $C_{4}$ rotational symmetry and share the tetragonal crystal system with $D_{4h}$ point group symmetry.
For this reason, we consider the undistorted topological DSM having a $D_{4h}$ point group symmetry as a representative model system.

\subsubsection*{Model and symmetry}
The general $4 \times 4$ Hamiltonian representation is
\begin{align} \label{eq:general_Hamiltonian}
 H(\mathbf k)  = \sum_{i=1}^{16} a_{i}  (\mathbf k) \Gamma_i.
\end{align}
The coefficient function $ a_{i}  (\mathbf k) $ are real functions 
and $\Gamma_i = s_j \sigma_k $ are $4 \times 4$ gamma matrices
where  $s_{j}$ and $\sigma_{k}$ are Pauli matrices for spin and orbital degrees of freedom
in the spin $(\uparrow, \downarrow)$ and the orbital $(1, 2)$ spaces, respectively.

The symmetry constraints can simplify the Hamiltonian's form in Eq.~(\ref{eq:general_Hamiltonian}).
Due to TRS and IS, the Hamiltonian satisfies the following equations:
\begin{align}\label{eq:Ham_symmetry_PT_1}
T H(\mathbf k) T^{-1} = H ( - \mathbf k), ~~~~~~
P H( \mathbf k) P^{-1} = H ( - \mathbf k), 
\end{align}
where $T = i s_y \hat K$ is the time-reversal operator ($\hat K$ is the complex conjugation operator)
and $P$ is the inversion operator.
Because the inversion does not flip the spin,
the inversion operator has orbital dependency only,
and it can be chosen as $P=  - \sigma_z$ for topological DSM
without loss of generality
\cite{wang2013three,yang2014classification}.
Then, due to TRS and IS, among sixteen $\Gamma_i$ matrices, only six $\Gamma_i$ matrices are allowed. They are 
$ \Gamma_0 = \mathbf{1}_{4 \times 4}$,
$ \Gamma_1 = \sigma_x s_z $,
$ \Gamma_2 = \sigma_y s_0 $,
$ \Gamma_3 = \sigma_x s_x $,
$ \Gamma_4 = \sigma_x s_y $, and
$ \Gamma_5 = \sigma_z s_0 $.
We set $a_0 (\mathbf k) = 0$ since it does not contribute to the formation of Dirac points \cite{wang2013three,yang2014classification}.

\begin{table}[b]
\centering
\begin{tabular}{   c |  c |  c  c  c c c c  c c c}
\hline \hline
& IR & $T$ & $P$ & $C_{4z}$ & $M_{xy}$ & $M_{yz}$   & $M_{xz}$   &  $M_{(110)}$ &  $M_{(1\bar{1}0)}$
\\
\hline \hline
$ \Gamma_0, \Gamma_5$ & $A_{1g}$  & $+$ & $+$ & $+$ & $+$ & $+$ & $+$ & $+$ & $+$ 
\\
$ \Gamma_4$ & $B_{1u}$ & $-$ & $-$ & $-$ & $-$ & $-$ & $-$ & $+$ & $+$ 
\\
$ \Gamma_3$ & $B_{2u}$  & $-$ & $-$ & $-$ & $-$ & $+$  & $+$ & $-$ & $-$ 
\\
$ (\Gamma_1, \Gamma_2)$  & $E_{u}$ & $(-,-)$ & $(-,-)$ & $(\Gamma_2,-\Gamma_1)$ & $(+,+)$ & $(-,+)$ & $(+,-)$ & $  (-\Gamma_2,-\Gamma_1)$ & $  (\Gamma_2,\Gamma_1)$
\\
\hline \hline
\end{tabular}
\caption{
\label{table:Gamma_condition_D4h}
Transformation properties of gamma matrices under symmetry operations.
Under an operation $O$, each gamma matrices satisfies the relation of $ O \Gamma_i O^{-1} = \pm \Gamma_j$.
In each entry, if $i=j$, the overall sign is written, otherwise the explicit form is given.
The gamma matrices are classified according to the irreducible representation (IR) of $D_{4h}$ point group.
$\Gamma_0$, $\Gamma_5$, $\Gamma_4$, and $\Gamma_3$ belong to the $A_{1g}$, $A_{1g}$, $B_{1u}$, and $B_{2u}$ irreducible representations, respectively.
$\Gamma_1$ and $\Gamma_2$ belong to the two-dimensional $E_{u}$ irreducible representation.
}
\end{table}

The $D_{4h}$ point group symmetry imposes more constraints on the Hamiltonian's form in Eq.~(\ref{eq:general_Hamiltonian}).
The generators of $D_{4h}$ point group can be chosen 
as inversion $P$, fourfold rotation about the $z$ axis $C_{4z}$, and twofold rotation about the $x$ axis $C_{2x}$.
Their matrix representations are chosen as
\begin{align}\label{eq:explict_C_4h}
P  =  - \sigma_z,~~~~~
C_{4z}  =  \exp ( - i \frac{\pi}{2}s_z - i \frac{\pi}{4} \sigma_z s_z ),~~~~~
C_{2x}  =   i \sigma_z s_x,
\end{align}
where we adopt the following basis set known to describe the low-energy effective Hamiltonian of Cd$_2$As$_3$\cite{wang2013three}.
\begin{equation}
\Ket{1, \uparrow}  = \Ket{P_{J=\frac{3}{2}} , 3/2},~~~~
\Ket{1,\downarrow} =\Ket{P_{J=\frac{3}{2}} , -3/2},~~~~
\Ket{2, \uparrow}  =\Ket{S_{J=\frac{1}{2}} , 1/2},~~~~
\Ket{2,\downarrow}  =\Ket{S_{J=\frac{1}{2}} , -1/2},
\end{equation}
where $J$ is the total angular momentum.
Other rotation and mirror symmetries are given by
$C_{2z} =  i \sigma_z s_z$,
$M_{xy} =   - i s_z$,
$M_{yz} =   - i s_x$,
$M_{zx} =  - i \sigma_z s_y$,
$M_{(1 1 0)}  =   i ( \sigma_z  s_x - s_y ) / \sqrt{2} $, and
$M_{(1 \bar 1 0)} = i ( \sigma_z  s_x + s_y ) / \sqrt{2} $.
The subscript in each mirror operator represents the corresponding mirror plane by using either Cartesian coordinates or Miller indices.
The group elements are derived in Sec.~S1 in Supplementary Information.
Due to this $D_{4h}$ symmetry, the Hamiltonian in Eq.~(\ref{eq:general_Hamiltonian}) satisfy
\begin{align}\label{eq:Ham_symmetry_D4h_01}
U H(\mathbf k) U^{-1}  = H (  S \mathbf k ),
\end{align}
where $U$ and $S$ are transformation matrices for an element of $D_{4h}$ group in the spin-orbital and momentum spaces, respectively.
For the group generators, the Hamiltonian in Eq.~(\ref{eq:general_Hamiltonian}) satisfies
\begin{align}
P H(\mathbf k) P^{-1} = H( - \mathbf k),~~~~~
C_{4z} H(\mathbf k) C_{4z}^{-1}  =  H( \mathcal{R}_{4z} \mathbf k),~~~~~
C_{2x} H(\mathbf k) C_{2x}^{-1}  =  H( \mathcal{R}_{2x} \mathbf k), 
\end{align}
where $\mathcal{R}_{4z} \mathbf k = (-k_y, k_x, k_z)$ and $\mathcal{R}_{2x} \mathbf k = (k_x, -k_y, -k_z)$.
Because the transformation properties of gamma matrices are given by Table~\ref{table:Gamma_condition_D4h},
Eq.~(\ref{eq:Ham_symmetry_D4h_01}) imposes constraints to each coefficient functions $a_i (\mathbf k)$, which is summarized in Table~\ref{table:a_function_additional_condition_D4h}.
Therefore, the general form of the Hamiltonian of DSM having $D_{4h}$ point group symmetry is obtained.

\begin{table}[b]
\centering

\begin{tabular}{   c |  c c  c  c c c c  c c }
\hline \hline
 & $T$ & $P$ & $C_{4z}$ & $M_{xy}$ & $M_{yz}$   & $M_{xz}$   &  $M_{(110)}$ &  $M_{(1\bar{1}0)}$  
\\
\hline \hline
$a_0(S \mathbf k),a_5(S \mathbf k)$ & $+$ & $+$ & $+$ & $+$ & $+$ & $+$ & $ +$ & $ +$ 
\\
$(a_1(S \mathbf k), a_2(S \mathbf k))$ & $(-,-)$ & $(-,-)$ & $(-a_2(\mathbf k),a_1(\mathbf k))$ & $(+,+)$ & $(-,+)$ & $(+,-)$ & $(- a_2(\mathbf k),- a_1(\mathbf k))$ & $ (a_2(\mathbf k),a_1(\mathbf k))$ 
\\
$a_3(S \mathbf k)$ & $-$ & $-$ & $-$ & $-$ & $+$  & $+$ & $-$ & $-$ 
\\
$a_4(S \mathbf k)$ & $-$ & $-$ & $-$ & $-$ & $-$ & $-$ & $+$ & $+$ 
\\
\hline \hline
\end{tabular}
\caption{
\label{table:a_function_additional_condition_D4h}
Symmetry constraints on $a_i(\mathbf k)$. They are determined by Eq.~(\ref{eq:Ham_symmetry_D4h_01}).
If the coefficient function is proportional to itself, $a_i (S\mathbf k) = \pm a_i (\mathbf k)$, the overall sign is denoted.
If not, the explicit form is denoted.
}
\end{table}

\subsubsection*{Lattice model}
\begin{figure}[ht]
\centering
\includegraphics[width=160mm]{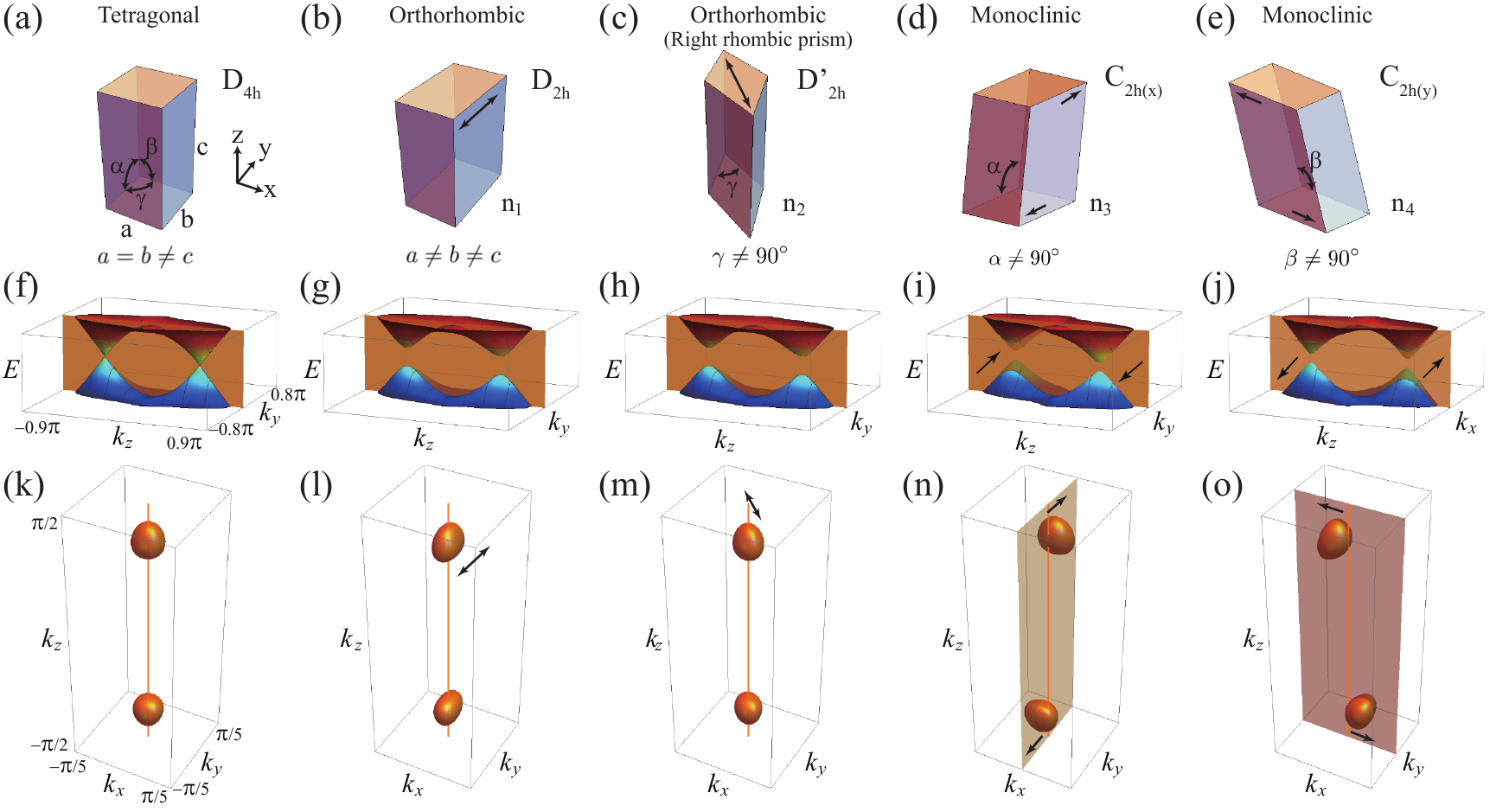} 
\caption{ 
\label{fig1:lattice_distortions}
\textbf{Crystal systems, band structures, and Fermi surfaces of Dirac semimetal (DSM) under various lattice distortions.}
\textbf{(a)} 
Undistorted DSM for comparison. It has a tetragonal lattice.
\textbf{(b-e)} 
Distorted crystal systems under
(b) $n_1$, (c) $n_2$, (d) $n_3$, and (e) $n_4$ type lattice distortions.
In (b) and (c), $n_1$ and $n_2$ type lattice distortions changes inplane lattice constants, which results in orthorhombic lattices.
In (d) and (e), $n_3$ and $n_4$ type lattice distortions change the $\alpha$ and $\beta$ angles, 
which results in monoclinic lattices.
\textbf{(f-j)} The corresponding 3D band structures.
In (f-i) [(j)], the band structures are plotted for the $k_y$-$k_z$ ($k_x$-$k_z$) plane and the orange planes are $k_y=0$ ($k_x=0$) plane.
\textbf{(k-o)} The corresponding Fermi surfaces.
In (l-o), all Fermi surfaces are distorted according to types of lattice distortions.
In (n) and (o), the Fermi surfaces are shifted as indicated by the black arrows.
Each vertical orange line indicates the $k_z$ axis. 
} 
\end{figure}

For concreteness, we construct an explicit lattice model that describes a class of Dirac semimetals such as Cd$_3$As$_2$ and Au$_2$Pb. The coefficient functions of Hamiltonian in Eq.~(\ref{eq:general_Hamiltonian}) are given by \cite{yang2014classification,wang2013three}
\begin{align} 
a_1 (\mathbf k)  & =  v \sin k_x, \label{eq:DSM_lattice_model_1} \\
a_2 (\mathbf k)  & =  v \sin k_y, \label{eq:DSM_lattice_model_2} \\
a_3 (\mathbf k)  & = ( \beta + \gamma ) \sin k_z (\cos k_y - \cos k_x), \label{eq:DSM_lattice_model_3} \\
a_4 (\mathbf k)  & = - ( \beta - \gamma ) (\sin k_z \sin k_x \sin k_y), \label{eq:DSM_lattice_model_4} \\
a_5 (\mathbf k)  & = M' - t_{xy}(\cos k_x + \cos k_y ) - t_z \cos k_z, \label{eq:DSM_lattice_model_5}
\end{align}
where $M'$, $t_{xy}$, $t_z$, $v$, $\beta$, and $\gamma$ are material-dependent parameters.
The energy eigenvalues are given by 
\begin{align} \label{eq:energy_eigenvalue}
E = \pm \abs{a(\mathbf k)}  = \pm \left ( \sum_{i=1}^5 a_{i}^2 (\mathbf k) \right )^{1/2}.
\end{align}
If $t_z > (M'- 2 t_{xy}) >0$, the Hamiltonian hosts a pair of Dirac points at $(0, 0, \pm k_0)$ as shown in Fig.~\ref{fig1:lattice_distortions}(a).
Here, $k_0$ is determined by $ M\rq{} - 2 t_{xy} - t_z \cos k_0 =0$.
These Dirac points are protected by the $C_{4z}$ symmetry\cite{yang2014classification}.
Due to the $C_{4z}$, the four bands on the $k_z$ axis can have different $C_{4z}$ eigenvalues, which lead to fourfold degenerate Dirac points.

\subsubsection*{Low-energy effective Hamiltonian}
Near the Dirac points $(0, 0, \pm k_0)$, the low-energy effective Hamiltonian takes the form of Dirac Hamiltonian, which is given by
\begin{align}\label{eq:D4h-low-energy-effective-Ham}
H_{\text{Dirac}}^{(\pm)} =  v k_x \Gamma_1 + v k_y  \Gamma_2 \pm v_z (k_z \mp k_0) \Gamma_5.
\end{align}
where $v_z = t_z k_0$.
The energy spectrum shows anisotropic energy-momentum dispersion, which is given by
\begin{align} \label{eq:energy_Dirac}
E = 
\pm \sqrt { v^2 (k_x ^2 +  k_y)^2 + v_z^2 (k_z \mp k_0)^2  }.
\end{align}

\subsection*{Distorted Dirac semimetal}
\subsubsection*{Symmetry-lowering distortions}
In the absence of lattice distortions, Cd$_3$As$_2$\cite{aggarwal_unconventional_2016,he_pressure-induced_2016,wang2016observation} and Au$_2$Pb\cite{xing_superconductivity_2016,yu2017fully,wu2019ground} share the same $D_{4h}$ point group symmetry and show no superconductivity.
However, both materials showed superconductivity after the structural phase transition under pressure or cooling, and the superconducting critical temperature increases with the pressure\cite{he_pressure-induced_2016, wu2019ground}.
At the high pressure, Cd$_3$As$_2$ becomes a monoclinic lattice having $C_{2h}$ point group symmetry\cite{he_pressure-induced_2016}
and Au$_2$Pb becomes an orthorhombic lattice having $D_{2h}$ point group symmetry\cite{xing_superconductivity_2016}.
Thus, IS is preserved even under lattice distortions.
In addition, the superconductivity appears under the small lattice distortions in the hydrostatic experiments\cite{he_pressure-induced_2016,wu2019ground}.
Therefore, we assume that both TRS and IS are preserved under lattice distortions and the effect of the lattice distortion can be implemented as a perturbation\cite{stoneham2001theory}.

We now classify the possible symmetry-lowering lattice distortions.
The form of the perturbation Hamiltonian for the lattice distortions is given by
\begin{equation}
H_{\text{pert}} = \sum_{i=0}^{5} d_{i}(\mathbf k) \Gamma_i,
\end{equation}
where $d_{i}(\mathbf k)$ is a real-valued function of momentum and $\Gamma_i$ is the gamma matrix.
Because $\Gamma_{1}$, $\Gamma_{2}$, $\Gamma_{3}$, and $\Gamma_{4}$ are odd under $T$ and $P$,
the coefficient functions $d_{1}(\mathbf k)$, $d_{2}(\mathbf k)$, $d_{3}(\mathbf k)$, and $d_{4}(\mathbf k)$ are odd functions with respect to $\mathbf k$.
Similarly, the coefficient functions $d_{0}(\mathbf k)$ and $d_{5}(\mathbf k)$ 
are even functions with respect to $\mathbf k$.
Thus, the allowed lattice distortion terms can be either 
$k^{\text{odd}} ~ \Gamma_{1,2,3,4}$
or 
$k^{\text{even}} ~ \Gamma_{0, 5}$ types.

Because we assume TRS and IS to remain under lattice distortions, the Hamiltonians for distorted and undistorted DSM have the same form of $H = \sum_{i} a_{i}(\mathbf k) \Gamma_i$.
The only difference between the two Hamiltonians is the transformation properties of the coefficient function $a_{i} (\mathbf k)$.
In the absence of lattice distortions, 
$a_{i} (\mathbf k)$
needs to satisfy all transformation properties under all symmetry operations of $D_{4h}$ point group
in Table~\ref{table:a_function_additional_condition_D4h}.
However, in the presence of lattice distortion,  $a_{i} (\mathbf k)$ only needs to satisfy the transformation properties under the remaining symmetry operations, so $a_{i}(\mathbf k)$ is less constrained.

\subsubsection*{Lattice Hamiltonian with lattice distortions}
To discuss the effect of lattice distortions explicitly, we introduce the possible symmetry-lowering lattice distortions in the lattice model in Eqs.~(\ref{eq:DSM_lattice_model_1}-\ref{eq:DSM_lattice_model_5}).
For weak lattice distortions, the lattice distortions are approximately proportional to $\sin k_i$ and $\cos k_i$ as only nearest neighbor hoppings are relevant.
Because we are interested in the Dirac physics near the Dirac points $(0, 0, \pm k_0)$,
we assume that $k_x/k_z \ll 0$ and $k_y/k_z \ll 0$, which implies that $\sin k_x$ and $\sin k_y$ are 
smaller than $\sin k_z$ and $\cos k_i$.
Hence, $\sin k_z$ and $\cos k_i$ are dominant momentum dependent terms in the leading order, and the allowed lattice distortions are either 
$\sin k_z ~ \Gamma_{1,2,3,4}$ or  $\cos k_i ~ \Gamma_{0, 5}$ types.
Because $\cos k_i ~ \Gamma_{0, 5}$ types are included in the trivial $A_{1g}$ class of $D_{4h}$ point group, they do no break any symmetry.
On the other hand, $\sin k_z ~ \Gamma_{1,2,3,4}$ types are included in $B_{1g}$, $B_{2g}$, and $E_g$, and they break the crystal symmetry properly, which are summarized in Table~\ref{table:classification_distortion_leading}.
Therefore, in the leading order, there are four types of symmetry-lowering lattice distortion, which are given by 
\begin{align} \label{eq:lattice_mom_dependent_1}
H_{\text{pert}} = \sin k_z (n_1 \Gamma_3 + n_2 \Gamma_4 + n_3 \Gamma_2 + n_4 \Gamma_1),
\end{align}
where $n_i$ is the strength of each lattice distortion.
For convenience, each lattice distortion is denoted as $n_i$ type lattice distortions in this work.
From now on, we will consider these four types of symmetry-lowering lattice distortions, and the possible higher-order terms are discussed in Sec.~S2 in Supplementary Information.

\begin{table}[t]
\centering
\begin{tabular}{ cccccccc}
\hline \hline
Type & Form & Remaining subgroup &  Essential elements & Material \\
\hline \hline
$n_1$ & $ \sin k_z \Gamma_3 $  & $D_{2h}$ & $C_{2z}, M_{xy}, M_{xz}, M_{yz} $   & Au$_2$Pb  \\
$n_2$ & $ \sin k_z \Gamma_4 $ & $D_{2h}'$  & $C_{2z}, M_{xy}, M_{(110)}, M_{(1\bar 1 0)} $    & Au$_2$Pb  \\
$n_3$ & $ \sin k_z \Gamma_2 $ & $C_{2h(x)}$ & $C_{2x}, M_{yz} $ & Cd$_3$As$_2$ \\
$n_4$ & $ \sin k_z \Gamma_1 $ & $C_{2h(y)}$ & $C_{2y}, M_{xz} $  &  Cd$_3$As$_2$ \\
\hline \hline
\end{tabular}
\caption{
\label{table:classification_distortion_leading}
Four types of symmetry-lowering lattice distortions are classified according to the irreducible representation of $D_{4h}$ point group.
$n_1$ and $n_2$ belong to the $B_{1g}$ and $B_{2g}$ irreducible representations of $D_{4h}$, respectively, while $n_3$ and $n_4$ belong to the two-dimensional $E_{g}$ irreducible representation.
For each lattice distortion, the matrix form, remaining essential group elements, and related material are listed.
}  
\end{table}

Therefore, the coefficient functions in Eq.~(\ref{eq:general_Hamiltonian}) are given by \begin{align} \label{eq:lattice_Hamiltonian_distorted}
a_1 (\mathbf k)  & =     v \sin k_x + n_4 \sin  k_z , \\
a_2 (\mathbf k)  & =     v \sin k_y + n_3 \sin  k_z ,  \nonumber  \\
a_3 (\mathbf k)  & =  ( \beta + \gamma )  (\cos k_y - \cos k_x) \sin k_z +  n_1 \sin  k_z  ,  \nonumber  \\
a_4 (\mathbf k)  & =  - ( \beta - \gamma ) (\sin k_x \sin k_y \sin k_z ) + n_2 \sin  k_z,  \nonumber  \\
a_5 (\mathbf k)  & =  M' - t_{xy}(\cos k_x + \cos k_y ) - t_z \cos k_z.  \nonumber 
\end{align}
Under lattice distortion, the fourfold rotation symmetry is broken.
Thus, the Dirac point is gapped, which can be seen from the energy eigenvalues on the $k_z$ axis, 
$ E = \pm 
\sqrt{ 
( n_1^2 + n_2^2 + n_3^2 + n_4^2) \sin^2 k_z +a_{5} (0,0,k_z)^2
}$.
Thus, the Dirac point is gapped unless $n_1^2+n_2^2 + n_3^2 + n_4^2 = 0$.
As a result of the gap-opening, the DSM becomes a 3D topological insulator because of the band inversion at the $\Gamma$ point\cite{yang2014classification, kobayashi2015topological}.
Counting all the parity eigenvalues for the time-reversal-invariant momenta (TRIM) points of the bulk Brillouin zone (BZ)\cite{fu2007topological,hasan_colloquium:_2010}
gives a nontrivial $\mathbb{Z}_2$ invariant.

\subsubsection*{The effect of lattice distortions}
The four types of symmetry-lowering lattice distortions in Eq.~(\ref{eq:lattice_mom_dependent_1}) are classified according to
the irreducible representation of $D_{4h}$ group. The symmetry-lowering lattice distortions break $D_{4h}$ point group symmetry into its subgroup symmetry, which is summarized in Table~\ref{table:classification_distortion_leading}.
The $n_1$ and $n_2$ type lattice distortions are included in the one-dimensional class $B_{1g}$ and $B_{2g}$, and break $D_{4h}$ point group symmetry into $D_{2h}$ and $D_{2h}'$, respectively.
The $n_3$ and $n_4$ type lattice distortions are included in the two-dimensional class $E_{u}$ and break $D_{4h}$ point group symmetry into $C_{2h}$.
Note that $n_2$ type lattice distortion is related to the $n_1$ type lattice distortion via  $\pi/4$ rotation,
while  $n_4$ type lattice distortion is related to the $n_3$ type lattice distortion via  $\pi/2$ rotation.

We investigate the explicit effects of the lattice distortions on the crystal systems and the Fermi surfaces using the lattice model in Eq.~(\ref{eq:lattice_Hamiltonian_distorted}).
Figure~\ref{fig1:lattice_distortions} shows the crystal structures, the 3D band structures, and Fermi surfaces under various lattice distortions.
Under $n_1$ type lattice distortion, the crystal system and Fermi surface are elongated along $x$ or $y$ direction, $C_{4z}$ symmetry is broken, the Dirac point is gapped, and the crystal system becomes orthorhombic [Fig.~\ref{fig1:lattice_distortions}(b,g)].
Similarly, under the $n_2$ type lattice distortion, the crystal system and Fermi surface are elongated along diagonal lines either $x=y$ or $x=-y$, $C_{4z}$ symmetry is broken, the Dirac point is gapped, and the crystal system becomes orthorhombic [Fig.~\ref{fig1:lattice_distortions}(c,h)].
We denote the symmetry point group of this right rhombic prism as $D_{2h}'$.
Under $n_3$ type lattice distortion, the crystal structure undergoes structural phase transition from tetragonal to monoclinic [Fig.~\ref{fig1:lattice_distortions}(d)].
Two Dirac points in the band structure are shifted oppositely along $k_y$ direction and the centers of each Fermi surfaces are also oppositely shifted along the same $k_y$ direction [Fig.~\ref{fig1:lattice_distortions}(h)].
Similar effects occur under $n_4$ type lattice distortion [Fig.~\ref{fig1:lattice_distortions}(e,j)] because $n_4$ type lattice distortion are related with the $n_3$ type lattice distortion via $\pi/2$ rotation.
The point groups of these distorted systems under $n_3$ and $n_4$ type lattice distortions are denoted as $C_{2h(x)}$ and $C_{2h(y)}$, respectively.
Therefore, the four types of symmetry-lowering lattice distortions explain the lattice distortions of Cd$_3$As$_2$ and Au$_2$Pb under pressure.

\subsubsection*{Low-energy effective Dirac Hamiltonian under lattice distortions}

Near the Dirac points $(0,0, \pm k_0)$, 
the coefficient functions of the low-energy effective Hamiltonian can be approximated as
\begin{align} 
a_1 (\mathbf k)  & =     v  k_x + n_4  \sin k_0 , \nonumber \\
a_2 (\mathbf k)  & =     v k_y + n_3 \sin k_0 ,  \nonumber  \\
a_3 (\mathbf k)  & =  ( \beta + \gamma )  \left (  \frac{k_x^2 - k_y^2}{2} \right )  \sin k_0+  n_1  \sin k_0  ,  \nonumber  \\
a_4 (\mathbf k)  & =  - ( \beta - \gamma )  k_x  k_y \sin k_0 + n_2  \sin k_0,  \nonumber  \\
a_5 (\mathbf k)  & =  \pm v_z (k_z \mp k_0) \sigma_z.  \nonumber
\end{align}
With this low-energy effective Hamiltonian, we show that the lattice distortion acts as a Dirac mass term and increases DOS at Fermi surface.
We assume that the Fermi level is slightly above the Dirac points in undistorted lattice, or near the bottom of the conduction band minima after gap-opening at the Dirac points.

For $n_1$ and $n_2$ type lattice distortions, the low-energy effective Hamiltonian is given by
\begin{align}\label{eq:Ham_Dirac_n1_n2}
H_{\text{Dirac}}^{(\pm)}
=v k_x \Gamma_1 + v k_y  \Gamma_2 \pm v_z (k_z \mp k_0) \Gamma_5 \pm n_1 \sin k_0 \Gamma_3 \pm n_2 \sin  k_0 \Gamma_4.
\end{align}
So, $n_1$ and $n_2$ type lattice distortion terms act as Dirac mass terms. The energy eigenvalue is given by
\begin{align} \label{eq:energy_eigenvalue_n1_n2}
E = \pm \sqrt { v^2 (k_x^2 + k_y^2) + v_z^2 (k_z \mp k_0)^{2}  + \abs{n}^2 \sin^2 k_0  },
\end{align}
where $\abs{n} = \sqrt{n_1^2 + n_2^2}$.
By the assumption of the total electron number conservation under a weak lattice distortion,
the lattice distortion dependent DOS at the Fermi surface is given by
\begin{align} \label{eq:DSM_DOS_01}
\text{DOS}(\abs{n}) = \frac{1}{\pi v^2 v_z}\mu_0 \sqrt{\mu_0^2 + \abs{n}^2 \sin^2 k_0},
\end{align}
which indicates that DOS at the Fermi level is enhanced under the lattice distortion.
Here, $\mu_0$ indicates the chemical potential of the undistorted lattice.
See the detailed derivations in Sec.~S2.4 in Supplementary Information.

Next, we consider the $n_3$ type lattice distortion.
The $n_3$ type lattice distortion shifts the gap minima along the $k_y$ direction from $(0, 0,   \pm k_0)$ to $(0, \pm k_y^{(0)},  \pm k_0)$ with $ k_y^{(0)} = - n_3 \sin k_0/v$.
Then, the low-energy effective Hamiltonian near the gap minima points $(0, \pm k_y^{(0)},  \pm k_0)$ is given by
\begin{align} 
H_{\text{Dirac}}^{(\pm)}
=  v k_x \Gamma_1 + v (k_y \mp k_y^{(0)})  \Gamma_2  \pm v_z (k_z \mp k_0) \Gamma_5  \pm  m \Gamma_3, \nonumber
\end{align}
where  $m =  - (\beta + \gamma) \frac{n_3^2 \sin^3 k_0 }{2 v^2 }$ is the Dirac mass term.
The energy eigenvalue is given by
\begin{align} \label{eq:energy_eigenvalue_n3}
E = \pm \sqrt { v^2 k_x^2  +  v^2 (k_y \mp k_y^{(0)} )^2 + v_z^2 (k_z \mp k_0)^2  + m^2  }.
\end{align}
Similar to $n_1$ and $n_2$ type lattice distortions, 
DOS at the Fermi surface are given by
\begin{align} \label{eq:DSM_DOS_02}
\text{DOS}(n_3) = \frac{1}{\pi v^2 v_z}\mu_0 \sqrt{ \mu_0^2 + m^2 },
\end{align}
which means that the DOS at the Fermi level is enhanced under $n_3$ type lattice distortion.
Similarly, for $n_4$ type lattice distortion, the low-energy effective Hamiltonian and DOS are easily calculated because $n_3$ and $n_4$ type lattice distortions are related via $\pi/2$ rotation.

\subsubsection*{Multiple symmetry-lowering lattice distortions\label{sec:multiple-symmety-lowering-distortions}}
So far, we have considered only one type of lattice distortions.
However, more than two types of lattice distortions can be turned on simultaneously.
In this case, the final subgroup symmetry determines the crystal system and its physical properties.
When both $n_1$ and $n_3$ types lattice distortions are turned on,
the remaining subgroup has $P$, $C_{2x}$, $M_{yz}$ symmetries.
This subgroup is the same point group of the distorted Dirac semimetal under single $n_3$ type lattice distortion.
In other words, under $n_3$ type lattice distortion, the addition of $n_1$ type lattice distortion is also allowed.
A similar argument can be applied to $n_2$ and $n_4$ types lattice distortions.
When both $n_1$ and $n_2$ type lattice distortions are turned on, the remaining symmetries are $P$, $C_{2z}$, $M_{xy}$ symmetries. We denote this point subgroup as $C_{2h(z)}$, and we will not consider this case seriously because there is no real material that corresponds to this case. 
Similarly, the other combinations such as $(n_2, n_3)$, $(n_1, n_4)$, $(n_3, n_4)$, $(n_1, n_2, n_3)$, $(n_1, n_2, n_4)$ break all crystal symmetries except the inversion, and hence these cases are not interested in this work.

\subsection*{\label{sec:superconductivity}Superconductivity}
\subsubsection*{\label{sec:BdG} BdG Hamiltonian}
To discuss the effects of lattice distortions on the superconductivity in doped DSM,
we construct the Bogoliubov-de Gennes (BdG) Hamiltonian within mean-field approximation while keeping TRS and the crystal symmetry\cite{alexandrov2003theory,bennemann2008superconductivity}. The BdG Hamiltonian is given by
\begin{align} \label{eq:BdG_Hamiltonian}
H_{\text{BdG}} 
& =  \int d \mathbf{k} 
\Psi_{\mathbf k}^{\dagger} \mathcal H (\mathbf k) \Psi_{\mathbf k},\\
\mathcal{H}(\mathbf{k} ) & =  [H (\mathbf{k}) - \mu] \tau_z + \Delta (\mathbf k) \tau_x,
\end{align}
where $\tau_i$ is the Pauli matrices in the Nambu space.
$\Delta (\mathbf k)$ and $\mu$ are a pairing potential and a chemical potential, respectively.
$H (\mathbf{k}) $ is the normal state Hamiltonian in Eq.~(\ref{eq:general_Hamiltonian}).
The basis is taken as
\begin{align}
\Psi_{\mathbf{k}}^{\dagger}
= 
(
c^{\dagger}_{1 \mathbf{k} \uparrow},
c^{\dagger}_{2 \mathbf{k} \uparrow},
c^{\dagger}_{1 \mathbf{k} \downarrow},
c^{\dagger}_{2 \mathbf{k} \downarrow},
c_{1 \mathbf{-k} \downarrow},
c_{2 \mathbf{-k} \downarrow},
-c_{1 \mathbf{-k} \uparrow},
-c_{2 \mathbf{-k} \uparrow} ).   
\end{align}
While the pairing mechanism of doped DSM is not known yet, we assume the following onsite density-density interaction as a superconducting pairing interaction\cite{fu2010odd,nakosai2012topological,kobayashi2015topological,hashimoto2016superconductivity}:
\begin{equation} \label{eq:density_interaction}
H_{\text{int}}(x) = - U [  n_1^1(x) + n_2^2(x) ] - 2V n_1(x) n_2(x),
\end{equation}
where
$ n_{i}(x)$ is the electron density operators for $i$th orbital ($i = 1, 2$). $U$ and $V$ are intra-orbital and inter-orbital interaction strengths, respectively,
and we assume that at least one of them is attractive and responsible for superconductivity.
Because the pairing interaction depends on the orbital and is local in $\mathbf x$, the mean-field pairing potential is orbital dependent but momentum independent: $ \Delta(\mathbf k) = \Delta $.

\subsubsection*{Symmetry of BdG Hamiltonian}
The BdG Hamiltonian in Eq.~(\ref{eq:BdG_Hamiltonian}) has time-reversal symmetry $T$, particle-hole symmetry $C$, and chiral symmetry $\Gamma$:
\begin{align}
T \mathcal{H} (\mathbf k ) T^{-1} & =  \mathcal{H} ( - \mathbf k ), \\
C\mathcal{H} (\mathbf k ) C^{-1} & =   - \mathcal{H} ( - \mathbf k ), \\
\Gamma \mathcal{H} (\mathbf k)  \Gamma^{-1} & = \mathcal{H} (\mathbf k),
\end{align}
where $T=is_y \sigma_0 \tau_0 \hat K$ and $ C= i s_y \sigma_0 \tau_y \hat K$ are time-reversal and particle-hole symmetry operators, respectively,
and $\Gamma = T C= s_0 \sigma_0 \tau_y$ is the chiral operator.
$\hat K$ is the complex conjugation operator.
Therefore, the BdG Hamiltonian belongs to in DIII class according to the classification table of topological insulator and superconductor\cite{schnyder2008classification}.

If the pairing potential satisfies
$ P \Delta(\mathbf k) P^{-1} = \eta_P \Delta (- \mathbf k) $,
the BdG Hamiltonian has the inversion symmetry:
\begin{align} \label{eq:BdG_inversion_symmetry}
\tilde P \mathcal{H} (\mathbf k ) \tilde P^{-1} & =  \mathcal{H} ( - \mathbf k ), ~~~
\text{with} ~~\tilde P = \text{diag}(P, \eta_P P),
\end{align}
where $P$ and $\tilde P$ are the inversion operators for the DSM and BdG Hamiltonians, respectively, and $\eta_P$ is the inversion parity.
If $\eta_P = 1$ ($\eta_P = -1$), the superconducting phase is an inversion-even-parity (inversion-odd-parity) superconductor.
For a single-orbital superconductor, $\tilde P$ is the identity operator, and an inversion-odd-parity (inversion-even-parity) pairing is equivalent to the spin-triplet (spin-singlet) pairing.
However, because of the spin-orbit coupling and multi-orbital band structure,
the pairings are more complex in our case.

From now on, we consider momentum independent pairing potentials, 
$ \Delta(\mathbf{k}) = \Delta$,
because we assume onsite pairing interaction as discussed in Eq.~(\ref{eq:density_interaction}).
In the absence of lattice distortions, the BdG Hamiltonian has $D_{4h}$ point group symmetry\cite{hashimoto2016superconductivity}.
If a pairing potential satisfies the transformation property of $ U \Delta s_y U^{T} s_y = \eta_U \Delta $ under a symmetry operation of $D_{4h}$ point symmetry group,
the BdG Hamiltonian satisfies the corresponding symmetry:
\begin{align}
\tilde U \mathcal{H} (\mathbf k) \tilde U^{-1}  = \mathcal{H}(S \mathbf k),
\end{align}
where $U$ is the symmetry operator in spin and orbital spaces, $\eta_U$ is a phase factor, and $\tilde U = \text{diag} ( U,  \eta_U  s_y U^* s_y)$
is the extended symmetry operator in the Nambu space.

For the generators of $D_{4h}$ point group, 
if the pairing potential satisfies
$C_{4z}  \Delta s_y  C_{4z}^{T} s_y =  \eta_{C_{4z}}  \Delta$ with $\eta_{C_{4z}} = e^{ i \pi r /2 }$ ($ r=0, \ldots, 3$) and
$C_{2x}  \Delta s_y  C_{2x}^{T} s_y =   \eta_{C_{2x}} \Delta$ with $\eta_{C_{2x}} = \pm 1$, then
the BdG Hamiltonian satisfies the corresponding rotation symmetry:
\begin{align}
\tilde C_{4z}  \mathcal{H} (\mathbf k) \tilde C_{4z}^{-1} & =   \mathcal{H} ( R_{4z}  \mathbf k), \\ 
\tilde C_{2x}  \mathcal{H} (\mathbf k) \tilde C_{2x}^{-1} & =   \mathcal{H} ( R_{2x} \mathbf k), 
\end{align}
where the extended symmetry operators are given by
$\tilde  C_{4z} = \text{diag} ( C_{4z},  \eta_{C_{4z}} s_y C_{4z}^* s_y )$ and $\tilde  C_{2x} = \text{diag} ( C_{2x},  \eta_{C_{2x}} s_y C_{2x}^* s_y )$.
If the pairing potential satisfies $M \Delta  s_y M^{T} s_y  = \eta_M  \Delta$ under a mirror operator $M$, the BdG Hamiltonian satisfies the corresponding mirror symmetry:
\begin{align}
\tilde{M} \mathcal{H} (\mathbf k_{\parallel}, \mathbf k_{\perp}) \tilde{M}^{-1}
= \mathcal{H} (\mathbf k_{\parallel}, - \mathbf k_{\perp}) , 
\end{align}
where $ \tilde M =\text{diag}~(M,  \eta_M  s_y M^* s_y) $ is a mirror operator for BdG Hamiltonian and $\mathbf k_{\parallel}$ ($\mathbf k_{\perp}$) is the momentum vector  parallel (perpendicular) to the mirror plane.
The $\eta_M$ is the mirror parity of the pairing potential under the mirror operation $M$.

In Table~\ref{table:pairing_potentials}, the transformation properties of all possible pairing potentials under the rotation and mirror operators are summarized. The details of each pairing potential will be discussed below.

\begin{table}[t]
\centering
\begin{tabular}{c c c c c c c c c c c c c c c c c c}
\hline 
\hline
Pairing  & $D_{4h}$ &$E$ & $P$ & $C_{4z}$ & $C_{2x}$
& $M_{xy}$ & $M_{yz}$ & $M_{zx}$ & $M_{110}$ & $M_{1 \bar 1 0}$
& fermion bilinear & matrix form \\
\hline
\hline
${\Delta}_{1}$ &$A_{1g}$ & 1 & 1& 1 & 1
& 1 & 1 & 1 & 1 & 1
& 
$c^{\dag}_{1\uparrow}c^{\dag}_{1\downarrow}+c^{\dag}_{2\uparrow}c^{\dag}_{2\downarrow} + h.c.$
& $\hat{I}$ 
\\
$\Delta_{1}'$ &$A_{1g}$ & 1 & 1& 1 & 1
& 1 & 1 & 1 & 1 & 1 
& 
$c^{\dag}_{1\uparrow}c^{\dag}_{1\downarrow}-c^{\dag}_{2\uparrow}c^{\dag}_{2\downarrow} + h.c.$
& $\sigma_z $ 
\\
${\Delta}_{2}$ &$B_{1u}$ & 1 & -1& -1 & 1  
& -1 & -1 &-1  & 1 & 1 
& $c^{\dag}_{1\uparrow}c^{\dag}_{2\uparrow}+c^{\dag}_{1\downarrow}c^{\dag}_{2\downarrow} + h.c.$
& $\sigma_{y}s_{y}$ 
\\
${\Delta}_{3}$ &$B_{2u}$ & 1 & -1& -1 & -1  
& -1 & 1 & 1 & -1 & -1 
& $ i (c^{\dag}_{1\uparrow}c^{\dag}_{2\uparrow}-c^{\dag}_{1\downarrow}c^{\dag}_{2\downarrow}) + h.c. $ 
& $\sigma_{y}s_{x}$ 
\\
${\Delta}_{41}$ &$E_{u}$ & 1  & -1& $\Delta_{42}$ & -1
& 1 & 1 & -1 &  $ \Delta_{42} $ &  -$ \Delta_{42} $
& $ c^{\dag}_{1\uparrow}c^{\dag}_{2\downarrow}-c^{\dag}_{1\downarrow}c^{\dag}_{2\uparrow} + h.c.$
& $\sigma_{x}$  
\\
${\Delta}_{42}$ &$E_{u}$ & 1  & -1& -$\Delta_{41}$ & 1
& 1 & -1 & 1 &  $ \Delta_{41} $ &  -$ \Delta_{41} $
& $i( c^{\dag}_{1\uparrow}c^{\dag}_{2\downarrow}+c^{\dag}_{1\downarrow}c^{\dag}_{2\uparrow}) + h.c.$
& $\sigma_{y}s_{z}$  
\\
\hline \hline
\end{tabular}
\caption{
\label{table:pairing_potentials}
The pairing potentials are classified according to the irreducible representation of $D_{4h}$ point group.
$\Delta_1$, $\Delta_1'$, $\Gamma_2$, and $\Gamma_3$ belong to the $A_{1g}$, $A_{1g}$, $B_{1u}$, and $B_{2u}$ irreducible representations, respectively.
$\Delta_{41}$ and $\Gamma_{42}$ belong to the two-dimensional $E_{u}$ irreducible representation.
The transformation properties of the pairing potentials are represented by $+1$ and $-1$ for even and odd parities.
For two-dimensional representation $E_u$, the explicit forms are listed.
} 
\end{table}

\begin{table}[b]
\centering
\begin{tabular}{ cccccccc  }
\hline
\hline
Pairing 
&  $D_{4h} $ 
&  $D_{2h} $
&  $D_{2h}'$
&  $C_{2h(z)} $ 
&  $C_{2h(x)} $  
&  $C_{2h(y)} $  
\\
\hline
\hline
$ {\Delta}_{1} $
& $ A_{1g} $ 
& $ A_{g} $ 
& $ A_{g} $ 
& $ A_{g} $ 
& $ A_{g} $ 
& $ A_{g} $ 
\\
$ \Delta_{1}' $
& $ A_{1g} $ 
& $ A_{g} $
& $ A_{g} $
& $ A_{g} $ 
& $ A_{g} $ 
& $ A_{g} $ 
\\
$ {\Delta}_{2} $ 
& $ B_{1u} $
& $ A_{u} $
& $ B_{1u} $
& $ A_{u} $
& $ A_{u} $
& $ A_{u} $
\\
$ {\Delta}_{3} $ 
& $ B_{2u} $
& $ B_{1u} $
& $ A_{u} $
& $ A_{u} $
& $ B_{u} $
& $ B_{u} $
\\
$ {\Delta}_{41} $
& $ E_{u} $ 
& $ B_{2u} $
& $ B_{3u}-B_{2u} $
& $ B_{u} $ 
& $ B_{u} $ 
& $ A_{u} $ 
\\
$ {\Delta}_{42} $
& $ E_{u} $ 
& $ B_{3u} $
& $ B_{3u}+B_{2u} $
& $ B_{u} $ 
& $ A_{u} $ 
& $ B_{u} $ 
\\
\hline 
\hline
\end{tabular}
\caption{ \label{table:pairing_with_lattice_distortion}
Pairing potentials classified according to the $D_{4h}$ point group are reclassified according to the irreducible representation of unbroken subgroup under the lattice distortions.
For $D_{2h}'$ group, $\Delta_{42}+\Delta_{41}$ and $\Delta_{42}-\Delta_{41}$ pairing potentials belong to in $ B_{3u}$ and $B_{2u}$ representations, respectively.
} 
\end{table}

\subsubsection*{Pairing potentials}
We now investigate the possible superconducting pairing potentials in the presence of lattice distortions.
Since we are considering multi-orbital superconductivity in the basis of  two spins and two orbitals, pairing potentials can be represented as a product of spin Pauli matrices and orbital Pauli matrices, which leads to sixteen matrices.
Among them, only six matrices are allowed
because of the fermion statistics ($\Delta s_y = s_y \Delta^T$).
We denote them as $\Delta_{1}$,
$\Delta_{1}'$, $\Delta_{2}$, $\Delta_{3}$, $\Delta_{41}$, and $\Delta_{42}$, whose forms and properties are listed in  Table~\ref{table:pairing_potentials}.
Due to Pauli's exclusion principle, the fermion bilinear form of each pairing potential shows antisymmetric property under the particle exchange. Because the pairing potential is momentum independent,
the spatial part is symmetric, while the spin-orbital part is antisymmetric under the particle exchange.
Thus, if the spin part is singlet, the orbital part is triplet, and vice versa.
Therefore, $\Delta_1$'s and $\Delta_{41}$ are the spin-singlet orbital-triplet pairings and $\Delta_2$, $\Delta_3$, and $\Delta_{42}$ are the spin-triplet orbital-singlet pairings as shown in the bilinear form in Table~\ref{table:pairing_potentials}.

Six pairing potentials can be classified according to the irreducible representations of the unbroken point group,
and the superconducting critical temperatures for the pairing potentials in the different classes are independent\cite{bennemann2008superconductivity,fu2010odd,nakosai2012topological,kobayashi2015topological,hashimoto2016superconductivity}.
In the absence of lattice distortions, the pairing potentials are classified according to the $D_{4h}$ group:
$\Delta_{1}$'s, $\Delta_2$, $\Delta_3$ and $\Delta_4$'s belong to $A_{1g}$, $B_{1u}$, $B_{2u}$ and $E_u$ irreducible representations, respectively, which are summarized in Table~\ref{table:pairing_potentials}.

The pairing potential belonging to a specific irreducible representation of the $D_{4h}$ group can be decomposed into a combination of different irreducible representations depending on the symmetry of the distorted lattice.
Some pairing potentials in the $D_{4h}$ group's individual representations can be included in the same representation and vice versa.
As an example, in the $D_{2h}$ case,
$(\Delta_{41}, \Delta_{42})$ belong to in the two-dimensional representation $E_u$ are separated into one-dimensional representations $B_{2u}$ and $B_{3u}$, respectively.
Similarly, for $D_{2h}'$ case, the linear combination of $\Delta_{41}$ and $\Delta_{42}$ potential belongs to in one-dimensional representations $B_{2u}$ and $B_{3u}$.
Because $D_{2h}'$ case is the $\pi/4$-rotated version of $D_{2h}$ case,
$\Delta_{41}+\Delta_{42}$ ($\Delta_{42}-\Delta_{41}$) is included in $B_{3u}$ ($B_{2u}$) class when $\Delta_{41}=\Delta_{42}$ ($\Delta_{41}=-\Delta_{42}$).
The reclassification of pairing potentials under various lattice distortions is summarized in Table~\ref{table:pairing_with_lattice_distortion}.

\begin{figure}[!t]
\centering
\includegraphics[width=160mm]{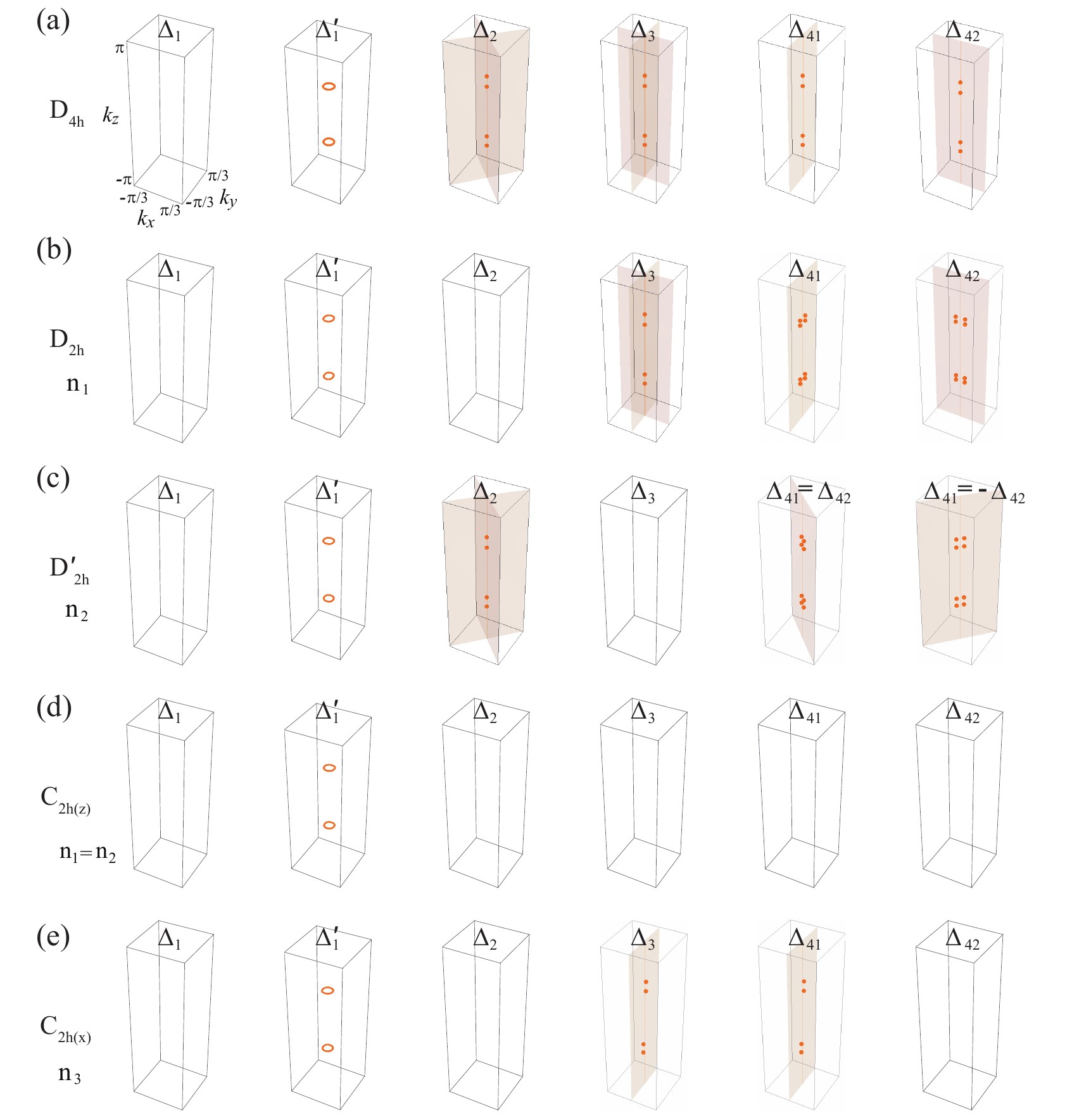} 
\caption{ 
\label{fig2:full_nodal_structure}
\textbf{Superconducting nodal structures for pairing potentials under lattice distortions.}
Nodal structures for \textbf{(a)}  $D_{4h}$, \textbf{(b)}  $D_{2h}$, \textbf{(c)}  $D_{2h}'$, \textbf{(d)}  $C_{2h(z)}$, and \textbf{(e)}  $C_{2h(x)}$ cases.
The orange point, line, and plane indicate nodal point and nodal line, and mirror plane ($M_{xz}$, $M_{yz}$, $M_{110}$, and  $M_{1 \bar 1 0}$), respectively.
In (a-e), the $\Delta_{1}$ phases are fully gapped and the $\Delta_{1}'$ phases have two nodal rings.
In (a,b,c,e), nodal points are located in the corresponding mirror planes.
In (c), $\Delta_{42}+\Delta_{41}$ and $\Delta_{42}-\Delta_{41}$ phases are considered instead of  $\Delta_{41}$ and $\Delta_{41}$ phases.
In (d), the system has no mirror symmetries and hence no nodal points.
These nodal structures are summarized in Table~\ref{table:nodal_structure}.
} 
\end{figure}

\begin{table}[b]
\centering
\begin{tabular}{  c l l l l l l   }
\hline \hline
&  $ {\Delta}_{1} $ 
&  $ \Delta_{1}' $ 
&  $ {\Delta}_{2} $  
&  $ {\Delta}_{3} $  
&  $ {\Delta}_{41} $ 
&  $ {\Delta}_{42} $
\\
\hline \hline
$D_{4h} $ 
& FG
& LN $^a$
& PN $^{b, c, d}$
& PN $^{b, c, d}$
& Acc. $^f$
& Acc. $^f$
\\
\hline \hline 
$D_{2h} $ 
& FG
& LN $^a$
& FG
& PN $^{c, d}$
& PN $^{c, e}$
& PN $^{c, e}$
\\
$D_{2h}'$
& FG
& LN $^a$
& PN $^{c, d}$
& FG
& PN $^{c, e}$
& PN $^{c, e}$
\\
\hline \hline
$C_{2h(z)} $
& FG
& LN $^a$
& FG
& FG
& FG 
& FG
\\
$C_{2h(x)} $ 
& FG
& LN $^a$
& FG
& PN $^{c, e}$
& PN $^{c, e}$
& FG 
\\
\hline \hline
\end{tabular}
\caption{ 
\label{table:nodal_structure}
Nodal structures of superconducting phases under lattice distortions.
FG, LN, and PN denote full gap, line node, and point node, respectively. 
$^a$ Topological line node protected by the chiral winding number ($W= \pm 2$ for each line node).
$^b$ Node protected by $C_{4z}$ symmetry.
$^c$ Topological point node protected by the mirror chiral winding number ($W_M = \pm 2$ for each point node).
$^d$ The nodal point is located on the $k_z$ axis.
$^e$ The nodal point is off the $k_z$ axis.
$^f$ Accidental point node.
}
\end{table}

\subsection*{\label{sec:nodal_structure}Superconducting nodal structure}
In this subsection, we classify the superconducting nodal structures under lattice distortions and study the symmetry and topology that guarantee the classified nodal structures.

Figure~\ref{fig2:full_nodal_structure} shows the typical nodal structures of superconducting phases of the doped DSM under lattice distortions.
There are three types of nodal structures: Full gap, point nodal, and line nodal structures, which are summarized in Table~\ref{table:nodal_structure}.
For $\Delta_{1}$ and $\Delta_{1}'$ superconducting phases, $\Delta_{1}$ phase is fully gapped and  $\Delta_{1}'$ phase has two nodal rings regardless of lattice distortions [Fig.~\ref{fig2:full_nodal_structure}(a-e)].
For $\Delta_{2}$ and $\Delta_{3}$ phases, nodal points exist at the intersections between the $k_z$ axis and the Fermi surfaces in the absence of lattice distortions
[Fig.~\ref{fig2:full_nodal_structure}(a)].
These points are known to be protected by $C_{4z}$ symmetry\cite{kobayashi2015topological,hashimoto2016superconductivity}.
Even under lattice distortions, if there is an unbroken mirror symmetry,
the topologically protected nodal points can exist and they are protected by the corresponding mirror symmetry [Fig.~\ref{fig2:full_nodal_structure}(b-e)].
For $\Delta_{41}$ and $\Delta_{42}$ phases,
there are accidental nodal points at the intersections between the $k_z$ axis and the Fermi surfaces in the absence of lattice distortions [Fig.~\ref{fig2:full_nodal_structure}(a)].
However, in the presence of lattice distortions, if there is an unbroken mirror symmetry, there can exist the topologically protected nodal points in the corresponding mirror plane [Fig.~\ref{fig2:full_nodal_structure}(b-e)]. 
Note that all nodal points under lattice distortions in Fig.~\ref{fig2:full_nodal_structure}(b-e) are protected by the topological mirror winding numbers, as discussed later.

We now analytically investigate the condition of nodal points in each superconducting phase.
Usually, nodal points can exist where the quasi-particle energy spectrum 
vanishes $\mathcal{E}(\mathbf k)=0$, which gives a set of equations for the momentum variables ($k_x, k_y, k_z$).
If the number of variables $N_{V}$ is greater than or equal to the number of independent equations $N_{E}$,
then nodal structures can exist. 
That is, $N_{E} \leq N_{V} = 3$ is the necessary condition for the existence of the nodes.
Moreover, if there is mirror symmetry, the necessary condition changes because the number of independent variables is reduced in the corresponding mirror plane.
That is, the necessary condition becomes $N_E \leq N_{V} = 2$.
If there is additional mirror symmetry, the necessary condition can be further reduced to $N_E \leq N_{V} = 1$ on the intersection of two mirror planes.

First, we consider $\Delta_{1}$ and $\Delta_{1}'$ superconducting phases.
The full gap structure of $\Delta_{1}$ phase is directly seen from the energy eigenvalues of 
\begin{align}
\mathcal{E}({\mathbf{k}})  = \pm \sqrt{(\abs{a} \pm \abs{\mu})^2 +  \Braket{\Delta_{1}}^2},
\end{align}
where $\abs a = \sqrt{ \sum_{i=1}^5 a_i(\mathbf k)^2}$.
Unless $\Braket{\Delta_{1}} = 0$, $\Delta_{1}$ phase is fully gapped.
For $\Delta_{1}'$ phase, the energy eigenvalues are given by 
\begin{align}
\mathcal{E}({\mathbf{k}})
= \pm
\sqrt{
    \abs{a}^2 + \mu^2 + \Braket{\Delta_{1}'}^2 
    \pm 2 
    \sqrt{
    \mu^2 \abs{a}^2  + \Braket{\Delta_{1}'}^2  \left( \abs{a}^2 - a_5(\mathbf k)^2 \right)}
}.
\end{align}
From $ \mathcal{E}(\mathbf k )=0$, one can obtain the following equations:
\begin{align}
\abs{a}^2   = \mu^2  + \Braket{\Delta_{1}'}^2, ~~~~ a_5(\mathbf k)  = 0.
\end{align}
Because the number of variable ($N_V = 3$) is larger than the number of equation ($N_E=2$), a one-dimensional solution can exist, which leads to the nodal lines.
Because this argument works regardless of the lattice distortions,
the nodal rings can exist for all cases in Fig.~\ref{fig2:full_nodal_structure}.
On the other hand, under some lattice distortions, a mixture of $\Delta_{1}$ and $\Delta_{1}'$ phases is allowed when $\Delta_{1}$ and $\Delta_{1}'$ are in the same representation as shown in Table~\ref{table:pairing_with_lattice_distortion}.
In such case, the gap structures have full gap (nodal lines) when $\abs{\Braket{\Delta_{1}}} > \abs{\Braket{\Delta_{1}'}}$   ($\abs{\Braket{\Delta_{1}}} < \abs{\Braket{\Delta_{1}'}}$) [Fig.~S2].
See the detailed calculation in Sec.~S3 in Supplementary Information.

Next, consider the $\Delta_{2}$ and $\Delta_{3}$ superconducting phases.
In the absence of lattice distortions, the nodal points in $\Delta_{2}$ and $\Delta_{3}$ phases are protected by $C_{4z}$ symmetry\cite{kobayashi2015topological,hashimoto2016superconductivity}.
On the other hand, under lattice distortions, a mirror symmetry can protect the nodal points that appear in Fig.~\ref{fig2:full_nodal_structure}(b,c,e).
For $\Delta_3$ phase, the energy eigenvalues are given by
\begin{equation}
\mathcal{E}(\mathbf{k}) = \pm 
\sqrt{
 \abs{a}^2 + \mu^2  +\Braket{\Delta_{3}}^2 
 \pm 2 \sqrt{ \mu^2 \abs{a}^2 +(a_3(\mathbf k)^2 +  a_5(\mathbf k)^2)  \Braket{\Delta_3}^2}
}.
\end{equation}
From $\mathcal{E}(\mathbf{k}) =0$, we get the following equations:
\begin{align} \label{eq:nodal_equation_Delta_3}
\abs{a}^2 = \mu^2 + \Braket{\Delta_{3}}^2,~~~
a_1(\mathbf k) = a_2(\mathbf k) =  a_4(\mathbf k) = 0.
\end{align}
Because $ N_E = 4 $ is larger than $ N_V = 3$, 
there seems to be no allowed nodal point.
However, mirror symmetries can allow nodal points.
For example, consider $D_{2h}$ point group with $M_{yz}$ and $M_{xz}$ mirror symmetries.
Under the $M_{yz}$ mirror operation, $a_1(\mathbf k)$ and $a_4(\mathbf k)$ are odd according to Table~\ref{table:a_function_additional_condition_D4h},
which gives $a_1(\mathbf k) = a_4 (\mathbf k)= 0$ at the mirror plane $(0, k_y, k_z)$.
Similarly, $M_{xz}$ mirror symmetry gives $a_2(\mathbf k) = a_4 (\mathbf k)= 0$ at the mirror plane $(k_x, 0, k_z)$. 
Thus, along the $k_z$ axis, $a_1(\mathbf k) = a_2(\mathbf k) = a_4(\mathbf k) = 0$ and Eq.~(\ref{eq:nodal_equation_Delta_3}) is reduced to
\begin{align} \label{eq:Delta_3_nodal_equation}
a_3^2 (k_z) +  a_5^2(k_z)  = \mu^2 + \Braket{\Delta_{3}}^2.
\end{align}
Because $N_E = 1$ is equal to $N_V = 1$, 
nodal points can exist as shown in Fig.~\ref{fig2:full_nodal_structure}(b).
However, when $M_{yz}$ and $M_{xz}$ mirror symmetries are broken,
the nodal points for the $\Delta_3$ phase are not protected as shown in Fig.~\ref{fig2:full_nodal_structure}(c,d).

Similarly, the nodal points in $\Delta_2$ phase can be understood using $M_{110}$ and $M_{1 \bar 1 0}$ mirror symmetries.
These mirror symmetries allow nodal points on the $k_z$ axis
in Fig.~\ref{fig2:full_nodal_structure}(c).
On the other hand, when $M_{110}$ and $M_{1 \bar 1 0}$ mirror symmetries are broken, the nodal points disappear as shown in Fig.~\ref{fig2:full_nodal_structure}(b,d).
For the $C_{2h(z)}$ case, a mixture of $\Delta_{2}$ and $\Delta_{3}$ phases is possible because $\Delta_{2}$ and $\Delta_{3}$ are included in the same $A_u$ representation.
However, there is no allowed nodal point as shown in Fig.~\ref{fig2:full_nodal_structure}(d) because there is no helpful mirror symmetry. 
See the details in Sec.~S3 in Supplementary Information.

Finally, consider $\Delta_{41}$ and $\Delta_{42}$ phases.
Without lattice distortions, there are accidental nodal points on the $k_z$ axis [Fig.~\ref{fig2:full_nodal_structure}(a)].
The existence of such nodal point is easily seen using four mirror symmetries $M_{xz}, M_{yz}, M_{110}$, and $M_{1 \bar 1 0}$.
These mirror symmetries force $a_i(\mathbf k) = 0$ for $i=1,\cdots,4$ on the $k_z$ axis according to Table~\ref{table:a_function_additional_condition_D4h}.
Then, the equations for nodal points are given by
\begin{align} 
\label{eq:41:condition}
\abs{a_5(k_z)}^2 =  \mu^2 + \Braket{\Delta_{41}}^2 + \Braket{\Delta_{42}}^2.
\end{align}
Because $N_E=N_V=1$, the nodal points exist.
Because the $\Delta_{41}$ and $\Delta_{42}$ pairing potentials included in $E_u$ representation of $D_{4h}$ point symmetry group, they break the $D_{4h}$ symmetry spontaneously to $D_{2h}$.
Hence, some of non-zero $a_i(\mathbf k)$ ($i=1\cdots4$) are spontaneously generated and the corresponding conditions are introduced, 
which makes the nodal points vanish.
Thus, these nodal points are accidental.
However, under lattice distortions, the nodal points can be protected by the unbroken mirror symmetry. For example, when the point group is $D_{2h}$ under the $n_1$ type lattice distortion, $\Delta_{41}$ and $\Delta_{42}$ are included in the different representations and thus we can consider each phase separately.
For $\Delta_{41}$ phase, $a_1(\mathbf k)=a_4(\mathbf k)=0$ on the mirror plane $(0, k_y, k_z)$ due to $M_{yz}$ symmetry.
Then, the equations for nodes are given by
\begin{align} 
a_2^2 (0, k_y, k_z) + a_5^2 (0, k_y, k_z) = \mu^2+ \Braket{\Delta_{41}}^2, ~~~
a_3 (0, k_y, k_z) = 0. 
\end{align}
Because $N_E=N_V=2$, there can exist nodal points [Fig.~\ref{fig2:full_nodal_structure}(b)].
For $\Delta_{42}$ phase, nodal points also can exist due to $M_{xz}$ mirror symmetry [Fig.~\ref{fig2:full_nodal_structure}(b)].
When the point group is $D_{2h'}$ under the $n_2$ type lattice distortion, nodal points can exist due to $M_{110}$ or $M_{1 \bar 1 0}$ mirror symmetries [Fig.~\ref{fig2:full_nodal_structure}(c)].
For $C_{2h(z)}$, a mixture of $\Delta_{41}$ and $\Delta_{42}$ phases is possible.
However, there is no allowed nodal point due to the lack of mirror symmetry [Fig.~\ref{fig2:full_nodal_structure}(d)].
When the point group is $C_{2h(x)}$ under the $n_3$ type lattice distortion, nodal points can exist due to $M_{yz}$ mirror symmetry [Fig.~\ref{fig2:full_nodal_structure}(e)].
See the detailed calculations in Sec.~S3 in Supplementary Information.

\begin{figure}[!t]
\centering
\includegraphics[width=160mm]{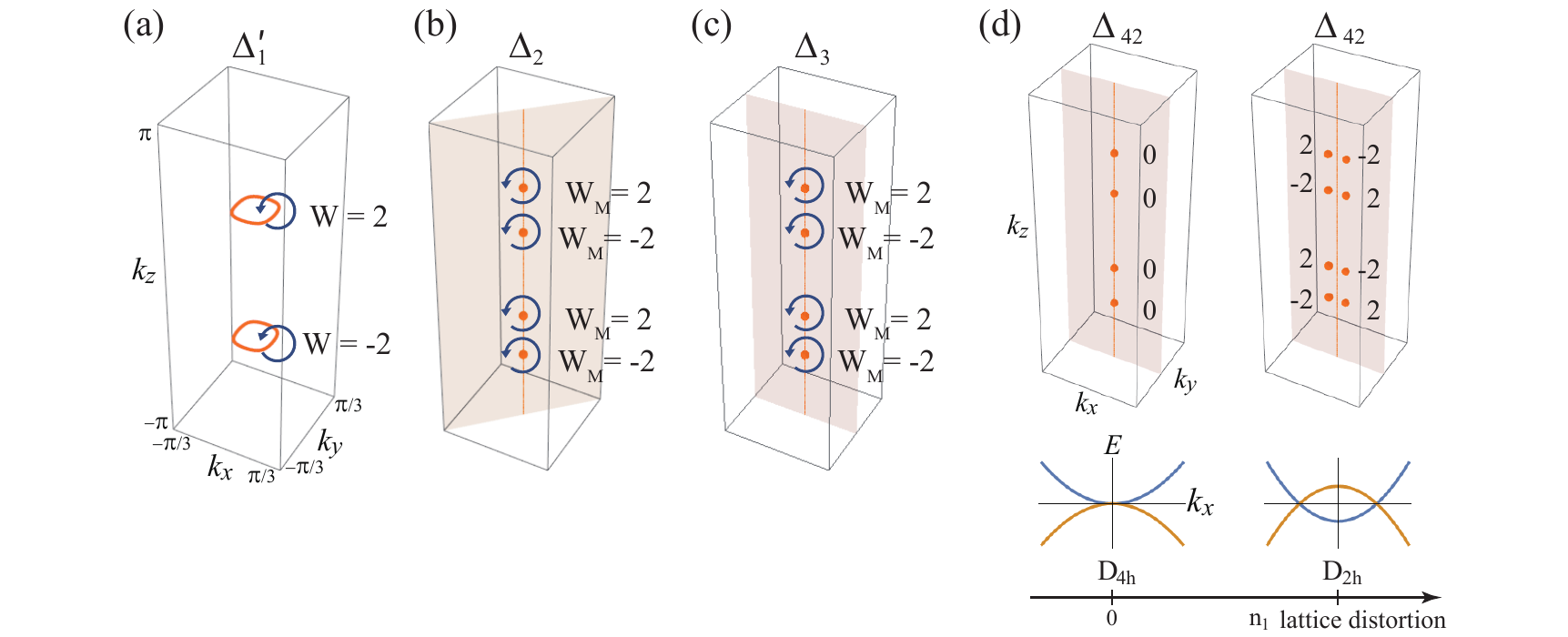} 
\caption{ 
\label{fig3:typical_winding_number}
\textbf{Topologically protected nodal structures and chiral winding numbers.}
The orange ring, point, plane, and vertical line indicate nodal ring, nodal point, mirror plane, and $k_z$ axis, respectively.
Each winding number is defined along each blue loop.
\textbf{(a)} The chiral winding numbers ($W=\pm 2$) protect nodal rings.
\textbf{(b,c)} The mirror chiral winding numbers ($W_M=\pm 2$) protect nodal points on the mirror planes.
\textbf{(d)} Evolution of nodal points in $\Delta_{42}$ phases and the corresponding mirror chiral winding number under the $n_1$ type lattice distortion. For clarity, the blue winding loops are omitted.
For $n_1=0$, nodal points with $W_M=0$ are located on $k_z$ axis.
These are fine-tuned accidental nodal points because $D_{4h}$ is spontaneously broken into $D_{2h}$ due to $\Delta_{42}$ pairing [see the main text below Eq.~ (\ref{eq:41:condition})].
As $n_1$ increases, the nodal points split into two nodal points with $W_M=\pm 2$.
The bottom plot shows the evolution of the energy dispersion along $k_x$ axis.
As $n_1$ increases, the blue (orange) band moves downward (upward), which results in two Dirac points.%
} 
\end{figure}

\subsection*{Stability of nodal structures}
There are two types of nodes in Table~\ref{table:nodal_structure},
which are symmetry-protected node and topologically-protected node.
In this subsection, we investigate the stability of them.

\subsubsection*{Chiral winding number}
Because of the chiral symmetry of the BdG Hamiltonian, 
the nodal lines can be protected by a chiral winding number \cite{schnyder2008classification,koshino2014topological,chiu2016classification}.
The chiral winding number is defined along a path $\mathcal{C}$ enclosing a singular point in the Brillouin zone as shown in Fig.~\ref{fig3:typical_winding_number}(a):
\begin{align}\label{eq:chiral_winding_number_def}
W
= \frac{1}{4 \pi i} \oint_{\mathcal{C}} \text{Tr} 
\left[
\Gamma \mathcal{H}^{-1} (\mathbf k) d  \mathcal{H}(\mathbf k)
\right],
\end{align}
where $\Gamma$ is the chiral operator.
As shown in Sec.~S4 in Supplementary Information,
the transformation property of the winding number under $P T$ symmetry is given by 
\begin{align}
W  =   -  \eta_{\Gamma, \tilde P T} W,
\end{align}
where the parity $\eta_{A,B} = \pm 1$  is determined by the relation $ AB = \eta_{A,B}  BA$.
For the inversion-even-parity (inversion-odd-parity) pairing potential, $\eta_{\Gamma, \tilde P T}$ is $-1$ ($+1$). 
Thus, the chiral winding number is zero for the inversion-odd-parity superconductor and only the inversion-even-parity superconducting phases having $\Delta_{1}$ and  $\Delta_{1}'$ pairing potentials can have a nontrivial chiral winding number.

\subsubsection*{$\Delta_{1}$ and $\Delta_{1}'$ phases}
Because $\Delta_{1}$ phase is fully gapped, the chiral winding number is zero.
On the other hand,
two nodal rings in $\Delta_{1}'$ phase are topologically protected by the chiral winding numbers. The calculated chiral winding numbers around the nodal rings are  $W=\pm2$ [Fig.~\ref{fig3:typical_winding_number}].
These chiral winding numbers do not change even under the lattice distortions because chiral winding number depends only on $T$, $C$, $P$, and $\Gamma$ symmetries. 
Thus, the topologically-protected  nodal rings in $\Delta_{1}'$ phase exist regardless of the lattice distortion [Fig.~\ref{fig2:full_nodal_structure}].

\subsubsection*{Mirror chiral winding number}
If there is mirror symmetry,
the BdG Hamiltonian commutes with the mirror symmetry operator in the mirror plane:
\begin{align}
[ \tilde{M},  \mathcal{H} (\mathbf k_{M})] = 0.
\end{align}
where $\tilde {M}$ is a mirror operator and $\mathbf k_{M}$ is the momentum vector located in the mirror plane.
Then, the BdG Hamiltonian can be block-diagonalized according to the mirror eigenvalues $\lambda = \pm i$.
Besides, if the mirror operator commutes with the chiral operator,
\begin{align} \label{eq:mirror_chiral}
	[ \Gamma, \tilde {M}] = 0 ,
\end{align}
the chiral operator also can be block diagonalized according to the same mirror eigenvalue.
Then, the winding number $W_{\lambda}$ in each mirror eigenvalue sector can be defined.
The condition in Eq.~(\ref{eq:mirror_chiral}) is satisfied 
only when the pairing potential is mirror even.
The reason is as follows:
In our convention,
the mirror operator for BdG Hamiltonian is defined as $\tilde {M} = \text{diag}[ M,  \eta_M s_y M^* s_y ] $
where $M$ and $s_y M^* s_y$ are mirror operators for electron part and hole part, respectively.
$\eta_M=\pm 1$ is the mirror parity of a pairing potential, which is given in Table~\ref{table:pairing_potentials}.
Because the mirror operator commutes with the time-reversal operator $[ T , M] = 0$,
all the mirror operator satisfies $s_y M^* s_y =  M$.
Then, $\tilde {M} = M \tau_0 $ ($\tilde {M} = M \tau_z$) for the mirror-even-parity (mirror-odd-parity) pairing potential.
Thus, only the mirror-even-parity superconducting phase satisfies the condition of Eq.~(\ref{eq:mirror_chiral}).

Furthermore, the mirror chiral winding number can be defined as $W_{M} = W_{i} - W_{-i}$,
where $W_{\lambda}$ is the chiral winding number for each block having a mirror eigenvalue $\lambda$.
The mirror chiral winding number $W_{M}$ can also be defined  
for a path $\mathcal{C}$ that encloses the Dirac point in the mirror plane 
as shown in Fig.~\ref{fig3:typical_winding_number}(b).
When the path $\mathcal{C}$ is parametrized by $\theta \in [0, 2 \pi )$, the mirror chiral winding number is given by\cite{hashimoto2016superconductivity,kobayashi2014topological}
\begin{align} \label{eq:mirror_chiral_winding_number}
W_{M}
= \frac{-1}{4 \pi} \oint_{\mathcal{C}} d \theta \text{Tr} 
\left[
\tilde {M} \Gamma \mathcal{H}^{-1} (\mathbf k(\theta)) d \mathcal{H}(\mathbf k(\theta))
\right].
\end{align}

\subsubsection*{$\Delta_2$ and $\Delta_3$ phases}
In the absence of lattice distortions, the $C_{4z}$ symmetry protects the nodal points by assigning different eigenvalues\cite{kobayashi2015topological,hashimoto2016superconductivity}.
The same nodal points are also topologically protected by the mirror chiral winding number in Eq.(\ref{eq:mirror_chiral_winding_number}) because the $\Delta_2$ and $\Delta_3$ pairing potentials are mirror-even.
For $\Delta_{3}$ pairing potential, which is mirror-even under $M_{xz}$ and $M_{yz}$, the calculated mirror chiral winding numbers around the nodal points are $\pm 2$ [Fig.~\ref{fig3:typical_winding_number}(c)].
Similarly, the nodal points in the $\Delta_{2}$ phase are topologically protected by $M_{110}$ and $M_{1 \bar 1 0 }$ mirror chiral winding numbers.

Even though $C_{4z}$ symmetry is broken under lattice distortions,
the mirror chiral winding number topologically protects the nodal points if the corresponding mirror symmetry is unbroken.
For example, consider $D_{2h}$ point group which has $M_{xz}$ and $M_{yz}$ mirror symmetries.
Among $\Delta_{2}$ and $\Delta_{3}$ pairings, $\Delta_{3}$ pairing is mirror even under $M_{xz}$ and $M_{yz}$.
Thus, the nodal points in the $\Delta_{3}$ phase are topologically protected by the corresponding mirror chiral winding numbers [Fig.~\ref{fig3:typical_winding_number}(c)].

Furthermore, the nodal points are positioned on the $k_z$ axis because  $C_{2z}$ symmetry gives an additional constraint as follows:
Let $W_M( \mathbf k)$ denote a mirror chiral winding number at $\mathbf k $.
Then, the mirror chiral winding number at $ C_{2z} \mathbf k $ is related
with that at $\mathbf k$ by
\begin{align} \label{eq:relation_mirror_C2z}
W_M( C_{2z}  \mathbf k) = \eta_{C_{2z}}W_M( \mathbf k),
\end{align}
where $\eta_{C_{2z}}$ is the parity of the pairing potential under $C_{2z}$ transformation.
The detail derivation is in Sec.~S4 in Supplementary Information.
Since $\eta_{C_{2z}}=1$ for $\Delta_{2}$ and $\Delta_{3}$, $W_M( \mathbf k) =W_M( C_{2z} \mathbf k)$,
which means that the mirror chiral winding numbers are the same for the two nodal points that are related by $C_{2z}$ rotation.
Now, let us assume that a nodal point on the $k_z$ axis in the absence of lattice distortions deviates from the $k_z$ axis under the $n_2$ type lattice distortion.
Due to the $C_{2z}$ symmetry, there exists another nodal point having the same mirror chiral winding number.
Thus, the total mirror winding number under lattice distortion becomes twice the original winding number, which is a contraction with the topological charge conservation.
Therefore, the nodal points should be located on the $k_z$ axis under the $n_2$ type lattice distortion.

A similar argument can be applied to the $D_{2h}'$ case having $M_{110}$ and $M_{1\bar 10}$ mirror symmetries.
The nodal points in the $\Delta_{2}$ phase is topologically protected by the  $M_{110}$ and $M_{1\bar 1 0}$ mirror chiral winding numbers
and the nodal points are located on the $k_z$ axis due to the $C_{2z}$ symmetry
[Figs.~\ref{fig2:full_nodal_structure}(c) and \ref{fig3:typical_winding_number}(b)].
For $C_{2h(x)}$ case, $M_{yz}$ is unbroken while $C_{2z}$ is broken.
Thus, nodal points on $M_{yz}$ plane in $\Delta_3$ phase are protected by the $M_{yz}$ mirror chiral winding number and can be deviated from $k_z$ axis due to the $C_{2z}$ symmetry breaking [Fig.~\ref{fig2:full_nodal_structure}(e)].

\subsubsection*{$\Delta_{41}$ and $\Delta_{42}$ phases}
In the absence of lattice distortions, the nodal points in each $\Delta_{41}$ and $\Delta_{42}$ phases [Fig.~\ref{fig2:full_nodal_structure}(a)] are accidental nodal points 
because a single phase, either $\Delta_{41}$ or $\Delta_{42}$ phase, would break the $D_{4h}$ point group symmetry spontaneously.
Only if we neglect such lattice symmetry breaking,
the accidental nodal points can be understood to be protected by the different eigenvalues of $C_{2z}$ and $s_z$ symmetry operators
(see the details in Sec.~3 in Supplementary Information).
Note that the existence of the accidental point nodes also can be verified via $C_{4z}$ symmetry \cite{kobayashi2015topological, hashimoto2016superconductivity}.
In the viewpoint of topological winding numbers, the mirror chiral winding numbers are zero in the absence of lattice distortions [Fig.~\ref{fig3:typical_winding_number}(d)] .
Due to the $C_{2z}$ symmetry, Eq.~(\ref{eq:relation_mirror_C2z}) gives 
\begin{align}\label{eq:Delta_1112_C2z_relation}
W_M( 0, 0, k_z) = - W_M( 0, 0, k_z),
\end{align}
which implies that $W_M=0$ on the $k_z$ axis.
Here, $\eta_{C_{2z}} =  -1$ is used for $\Delta_{41}$ and $\Delta_{42}$.
Thus, the nodal points are not topologically protected for $D_{4h}$ case.

\begin{figure}[!t]
\centering
\includegraphics[width=\linewidth]{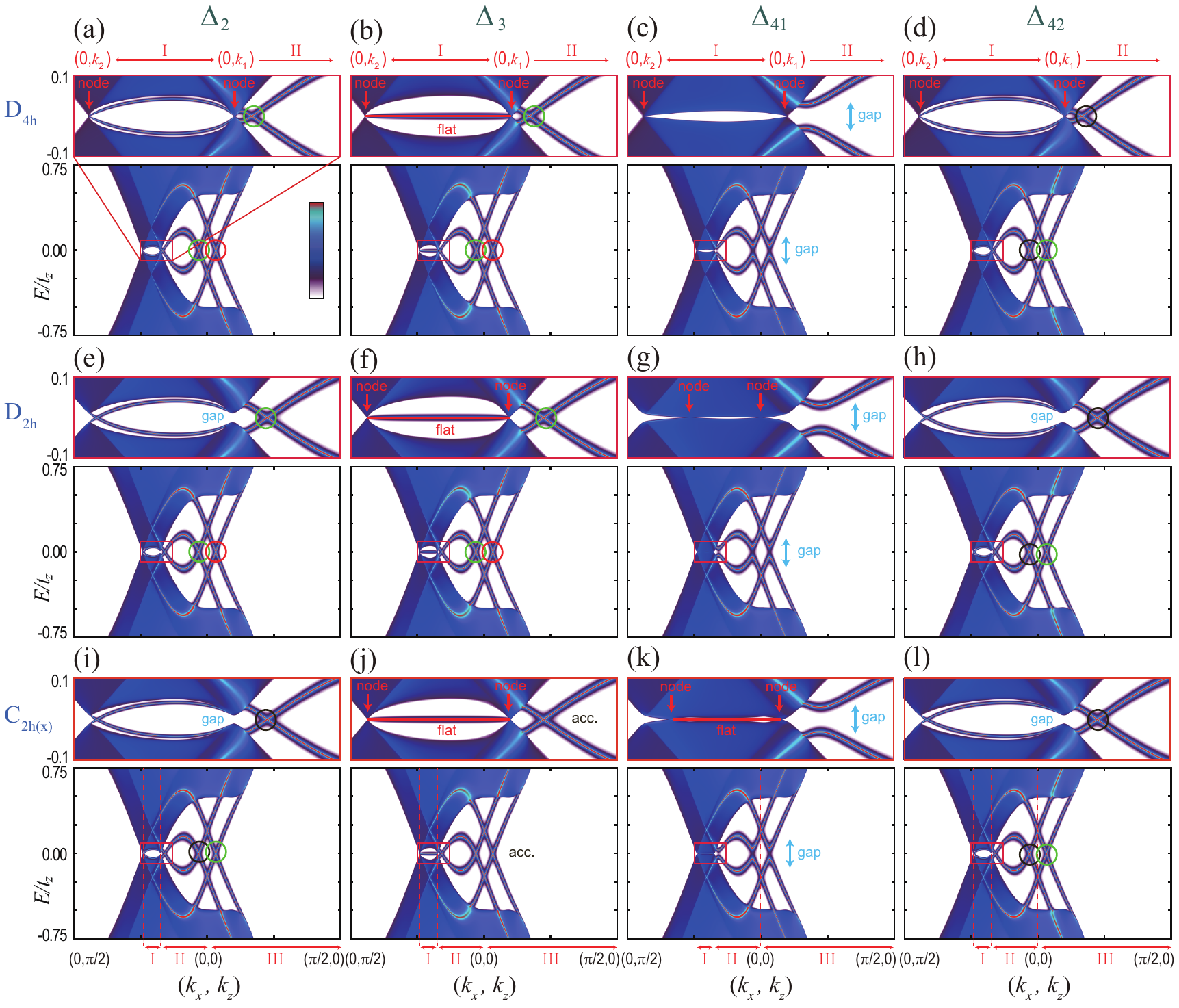} 
\caption{ 
\label{fig4:surface_spectrum}
\textbf{Surface band structures of superconducting phases under distortions.}
Surface band structures on the $(010)$ surface for 
\textbf{(a-d)} $D_{4h}$, \textbf{(e-h)} $D_{2h}$, \textbf{(i-l)} $C_{2h(z)}$ and \textbf{(m-p)} $C_{2h(x)}$.
In each panel, the upper figure indicates the close-up view of the band structure near $E=0$ corresponding to the red box in the lower figure.
The red vertical arrows indicate the nodal points of the bulk superconducting states.
In the insets of (e,h,i,l), the bulk states are gapped.
The cyan vertical arrows indicate the gapped surface states.
In (b,f,j,k), red horizontal lines show the surface flat bands.
The nature of gapless surface state (GSS) is distinguished by the colored circle:
Red ones in (a,b,e,f), green ones in (a,b,d,e,f,h,i,l), and black ones in (d,h,i,l) indicate GSS's protected by mirror Chern numbers, zero-dimensional topological numbers, and mirror eigenvalues, respectively.
In (j), GSS's are accidental.
The details are in Table \ref{table:summary_Majorana_fermion} and in the main text.
Region I, II, and III are $(0,k_1)$-$(0,k_2)$, $(k_2, 0)$-$(0,0)$, $(0,0)$-$(\pi/2,0)$, respectively, where $k_1$ and $k_2$  ($k_1>k_2>0$) indicate two intersecting points between the upper Fermi surface and the $k_z$ axis.
} 
\end{figure}

However, under lattice distortions, nodal points can be topologically protected by the mirror chiral winding number.
Let us consider the $D_{2h}$ point group under the $n_1$ type lattice distortion.
Since $\Delta_{41}$ and $\Delta_{42}$ pairings are mirror-even under $M_{yz}$ and $M_{xz}$ operations, the corresponding mirror chiral winding number protects nodal points in each mirror plane [Fig.~\ref{fig2:full_nodal_structure}(b)].
The calculated mirror chiral winding numbers are $W_M=\pm2$ [Fig.~\ref{fig3:typical_winding_number}(d)].
Note that the nodal points are off the $k_z$ axis
and the calculated mirror chiral winding numbers satisfy Eq.~(\ref{eq:Delta_1112_C2z_relation}).
For the $D_{2h}'$ case, $M_{110}$ and $M_{1 \bar 1 0}$ mirror chiral winding numbers ($W_M = \pm 2$) protect nodal points in corresponding mirror planes for the superconducting phases having $\Delta_{41} \pm \Delta_{42}$ pairing potentials [Fig.~\ref{fig2:full_nodal_structure}(c)].
For the $C_{2h(z)}$ case, all the relevant mirror symmetries are broken
and hence there are no topologically-protected nodal points [Fig.~\ref{fig2:full_nodal_structure}(d)].
For the $C_{2h(x)}$ case, there are unbroken $M_{yz}$ and $C_{2x}$.
Thus, $M_{yz}$ mirror chiral winding numbers ($W_M=\pm 2$) protect the nodal points but the nodal points need not to be located symmetrically with respect to the $k_z$ axis [Fig.~\ref{fig2:full_nodal_structure}(e)].
These nodal points in the $C_{2h(x)}$ case can be understood from the nodal points in the $D_{2h}$ case:
Among four nodal points in the $D_{2h}$ case, two nodal points are pair-annihilated, and only two nodal points survive in the $C_{2h(x)}$ case.

Finally, we discuss a gap structure change of $\Delta_{42}$ phase
under $n_1$ type lattice distortion [Fig.~\ref{fig3:typical_winding_number}(d)].
When $n_1=0$, each nodal points has $W_M=0$ and a quadratic energy-momentum dispersion relation along the $k_x$.
With the increasing lattice distortion, nodal points with $W_M=\pm 2$ are created pairwise from a nodal point with $W_M=0$, and linear energy-momentum dispersion relation for all three momentum directions appears.
Similar gap structure changes occur under the other lattice distortions.

%
%
%

\begin{table}[t]
\centering
\begin{tabular}{ |c|c|c|c|c|c|c|c|c|c|c|c|c| }
\hline
Pairing & \multicolumn{3}{c|}{ $\Delta_{2}$ } & \multicolumn{3}{c|}{ $\Delta_{3}$ } & \multicolumn{3}{c|}{ $\Delta_{41}$ } & \multicolumn{3}{c|}{ $\Delta_{42}$ }
\\
\hline
Region &  I & II & III  &  I & II & III  &  I & II & III  &  I & II & III
\\
\hline
$D_{4h} $ & $M_{yz}$ & $CM_{xy}$ & $C_M$  & $W_M$ & $CM_{xy}$ & $C_M$ & n/a & n/a & n/a & $M_{yz}$ & $M_{yz}$ & $\Gamma M_{yz}$
\\
\hline 
$ D_{2h} $ & $M_{yz}$ & $CM_{xy}$ & $C_M$ & $W_M$ & $CM_{xy}$ & $C_M$ & n/a & n/a & n/a & $M_{yz}$ & $M_{yz}$ & $\Gamma M_{yz}$
\\
\hline 
$C_{2h(x)} $ & $M_{yz}$ & $M_{yz}$ & $\Gamma M_{yz}$ & $W_M$ & Acc. & Acc. & $W_M$ & n/a & Acc. & $M_{yz}$ & $M_{yz}$ & $\Gamma M_{yz}$
\\
\hline 
\end{tabular}
\caption{ 
\label{table:summary_Majorana_fermion}
Gapless surface Andreev bound state (SABS) on $(010)$ surface.
The entry is either a topological number or a symmetry operator which protects corresponding gapless surface states.
Region I, II, and III are defined in Fig.~\ref{fig4:surface_spectrum}.
$W_M$ is a mirror chiral winding number that protects the flat SABS between nodal points.
$C_M$ is a mirror Chern number that protects the gapless SABS in $M_{xy}$ plane.
$\Gamma M_{yz}$ and $CM_{xy}$ indicate the symmetry operators which protect   gapless SABS using the corresponding zero-dimensional topological number.
$M_{yz}$ and $M_{xy}$ indicate the symmetry operators
which protect the gapless SABS protected by the corresponding mirror eigenvalues.
Acc. indicates an accidental gapless state.
n/a means that there is no gapless state.
}
\end{table}

\subsection*{Surface spectrum}
Surface Andreev bound state (SABS) in superconducting phases of the topological DSM have been studied in the absence of lattice distortion. \cite{hashimoto2016superconductivity}.
In this subsection, we systematically investigate SABS in superconducting phases under lattice distortions.
There are four types of gapless surface Majorana states under lattice distortions. Three types are topologically protected by mirror chiral winding, mirror Chern, and zero-dimensional winding numbers. The fourth type is protected by mirror symmetry and corresponding eigenvalues.

Using the Möbius transformation based method\cite{umerski1997closed}, we calculate the surface band structures.
Figure~\ref{fig4:surface_spectrum} shows the numerically obtained surface spectra for $(010)$ surface in various superconducting phases under lattice distortions.
For $\Delta_{1}$ and $\Delta_{1}'$ phases, there is no SABS;
$\Delta_{1}$ phase is fully gapped and topologically trivial, and 
$\Delta_{1}'$ phase has two nodal lines having opposite chiral winding numbers as shown in Fig.~\ref{fig3:typical_winding_number}(a),
which does not have protected SABS because of the positions and shapes of two nodes in momentum space.
On the other hand, $\Delta_{2}$, $\Delta_{3}$, $\Delta_{41}$, and $\Delta_{42}$ have various types of SABS [Fig.~\ref{fig4:surface_spectrum}], which are summarized in Table~\ref{table:summary_Majorana_fermion}.

Without loss of generality, we will focus on the $(010)$ surface
and the surface Brillouin zone $(k_{x}, k_{z})$.
A similar analysis for the $(010)$ surface can be easily applied to the other surfaces such as $(100), (110)$ planes, because the results for the other plane only depend on the mirror symmetries and the transformation properties of the pairing potentials under the unbroken symmetries.
For convenience, we consider the surface states in the three regions:
Region I, II, and III, which are $(0,k_1)$-$(0,k_2)$, $(k_2, 0)$-$(0,0)$, $(0,0)$-$(\pi/2,0)$, respectively.
Here, $k_1$ and $k_2$  ($k_1>k_2>0$) indicate two intersecting points between the upper Fermi surface and the $k_z$ axis.

First, we consider the flat SABS in the Region I, 
which is topologically protected by the nontrivial mirror chiral winding number in Eq.~(\ref{eq:mirror_chiral_winding_number}).
For example, let us consider $\Delta_{3}$ phase and $M_{yz}$ mirror symmetry.
For $D_{4h}$, $D_{2h}$, and $C_{2h(x)}$ cases,
$M_{yz}$ mirror is unbroken and $\Delta_{3}$ has odd parity under $M_{yz}$,
which leads to the opposite mirror chiral winding numbers ($W_M=\pm2$) for two nodal points near the upper Fermi sphere as shown in Fig.~\ref{fig3:typical_winding_number}(c).
Then, there exists a flat SABS on $(010)$ surface as shown in Fig.~\ref{fig4:surface_spectrum}(b,f,j).
To understand such SABS on $(010)$ surface, 
the mirror winding number $W_{M}(k_z)$ along the mirror invariant $k_z$ axis is defined as\cite{hashimoto2016superconductivity}
\begin{align} 
W_{M}(k_z)
= \frac{-1}{4 \pi i } \int_{-\pi}^{\pi} d k_y \text{Tr} 
\left[
\tilde {M} \Gamma \mathcal{H}^{-1}  d_{k_y} \mathcal{H}
\right],
\end{align}
which is nontrivial between nodal points.
Therefore, between the nodal points, there exists a flat SABS.
Similarly, for $\Delta_{41}$ phase, $M_{yz}$ mirror symmetry gives nontrivial mirror chiral winding numbers, which guarantees the existence of the zero-energy flat SABS in the Region I on $(010)$ surface [Fig.~\ref{fig4:surface_spectrum}(k)].
Note that, under the $n_3$ type lattice distortion, the mixture of $\Delta_{3}$ and $\Delta_{41}$ phases are allowed. But the flat SABS is still present due to the $M_{yz}$ mirror chiral winding number.

Second, we consider the gapless SABS protected by the mirror Chern number $C_M$.
The topological mirror superconducting phases\cite{kobayashi2015topological,zhang2013topological} are allowed for $\Delta_{2}$ and $\Delta_{3}$ phases because $\Delta_{2}$ and $\Delta_{3}$ pairing potentials are $M_{xy}$ mirror-odd and the corresponding mirror Chern numbers for each mirror eigenvalue block are nontrivial.
Under the $n_1$ ($n_2$) type lattice distortion, $\Delta_{2}$ ($\Delta_{3}$) phase is fully gapped, and the  mirror Chern number defined in $M_{xy}$ plane is nontrivial ($C_{M} = \pm 2$),
which leads to a topologically-protected Majorana states on $M_{xy}$ plane. For example, see the surface spectra in the Region III in Fig.~\ref{fig4:surface_spectrum}(e).

Third, we consider the gapless SABS protected by the zero-dimensional topological number.
Since $\Delta_{2}$ and $\Delta_{42}$ pairings are odd under $M_{yz}$,
a zero-dimensional topological number $\rho(k_x)$ can be defined using $\Gamma M_{yz}$\cite{kobayashi2015topological,hashimoto2016superconductivity}. Then, the zero-dimensional topological number protects the gapless state in the Region III.
See the surface spectra at the Region III in Fig.~\ref{fig4:surface_spectrum}(d, h, i, l) and Table~\ref{table:summary_Majorana_fermion}.
Similarly, $\Delta_{2}$ and $\Delta_{3}$ pairings are odd under $M_{yz}$,
a zero-dimensional topological number $\rho(k_z)$ is defined using $C M_{yz}$\cite{kobayashi2015topological,hashimoto2016superconductivity}, which protects the gapless states in the Region II for $D_{4h}$ and $D_{2h}$ cases.
See the surface spectra at the Region II in Fig.~\ref{fig4:surface_spectrum}(a, b, e, f) and Table~\ref{table:summary_Majorana_fermion}.

Fourth, we consider the gapless SABS protected by mirror eigenvalues.
If the pairing potential has an odd parity under the mirror operation,
the mirror eigenvalues for the electron and hole bands are different, which protects the band crossing of surface states\cite{kobayashi2015topological,hashimoto2016superconductivity}.
For example, consider $\Delta_{2}$ phase and $M_{yz}$ symmetry.
Because $(k_x, k_y, k_z) \rightarrow (-k_x, k_y, k_z)$ under $M_{yz}$, the mirror eigenvalues are properly defined on the $k_z$ axis.
Moreover, $\Delta_{2}$ pairing has odd parity under $M_{yz}$ symmetry.
Hence, the different mirror eigenvalues protect the gapless states in the Region I.
See Fig.~\ref{fig4:surface_spectrum}(a, e, i).
Similarly, $\Delta_{42}$ phases has odd parity under $M_{yz}$,
which protects the gapless states in the Region I and II.
See Fig.~\ref{fig4:surface_spectrum}(d, h, l).

In summary, we find the various types of surface states depending on the pairing potentials and lattice distortions.
Even under the lattice distortions, most of the inversion-odd-parity superconducting phases have gapless SABS, which may be observed as zero bias conductance peak (ZBCP) in experiments.

\subsection*{\label{sec:Tc_and_phase_map}Superconducting critical temperature and phase diagram}
In this subsection, we study superconducting critical temperatures and their enhancements under lattice distortions. We also investigate the phase diagram for the various superconducting phases under lattice distortion.

In the weak-coupling limit, the superconducting critical temperature $T_c$ can be calculated by solving the linearized gap equation and a phase diagram for various pairing potentials is obtained by comparing the critical temperatures\cite{alexandrov2003theory,bennemann2008superconductivity,fu2010odd,nakosai2012topological,hashimoto2016superconductivity}.
The linearized gap equation can be expressed using the pairing susceptibility\cite{alexandrov2003theory,bennemann2008superconductivity,fu2010odd,nakosai2012topological,hashimoto2016superconductivity}.
The pairing susceptibility $\chi_i$ for each pairing potential $\Delta_i$ is given by
\begin{align}\label{eq:def_susceptibility}
\chi_i(T)
= - \frac{1}{\beta} \sum_{\omega_n} \sum_{\mathbf k} \text{Tr} [ (\Delta_i  \tau_x) G_0(\mathbf k) (\Delta_i  \tau_x) G_0(\mathbf k)].
\end{align}
Here, $\beta = 1/(k_B T)$ is the inverse temperature, $k_B$ is the Boltzmann constant,
$\omega_n$ is the Matsubara frequency, and
$\Delta_i$ is the matrix representation of a pairing potential listed in Table~\ref{table:pairing_potentials}.
$G_0 (\mathbf k) = \frac{ \mathcal P_{\mathbf k}}{i \omega_n - \varepsilon_{\mathbf k}} $ is the single-particle Green's function of the normal state and $\mathcal P_{\mathbf k} \equiv \sum_{m=1,2}  \Ket{\phi_{m,\mathbf k} }\Bra{\phi_{m,\mathbf k }}$ is the projection operator onto the two degenerate Bloch states in the conduction bands.
Here, $\varepsilon_{\mathbf k}  =\abs{a(\mathbf k )} - \mu $.
Then, the superconducting susceptibility has the following generic form:
\begin{align}
\chi_i(T) = 
\int \frac{ d ^3 \mathbf k }{(2\pi)^3}
f_i(\mathbf k) \frac{ \tanh ( \beta \varepsilon_{\mathbf{k}}/2) }{ 2 \varepsilon_{\mathbf{k}} },
\end{align}
where 
$ f_{i} (\mathbf k) $ is momentum dependent form factor.
The explicit expressions for form factors $f_i ( \mathbf k) $ are given in Sec.~S5 in Supplementary Information.

With these susceptibilities, we now solve the linearized gap equation.
The linearized gap equations are obtained by minimizing the mean-field free energy in the weak coupling limit.
Since superconducting critical temperatures with pairing potentials in the same classes are not independent,
the $\Delta_{i}$'s in the same class can appear in the same linearized gap equation.

\begin{figure}[!t]
\centering
\includegraphics[width=120mm]{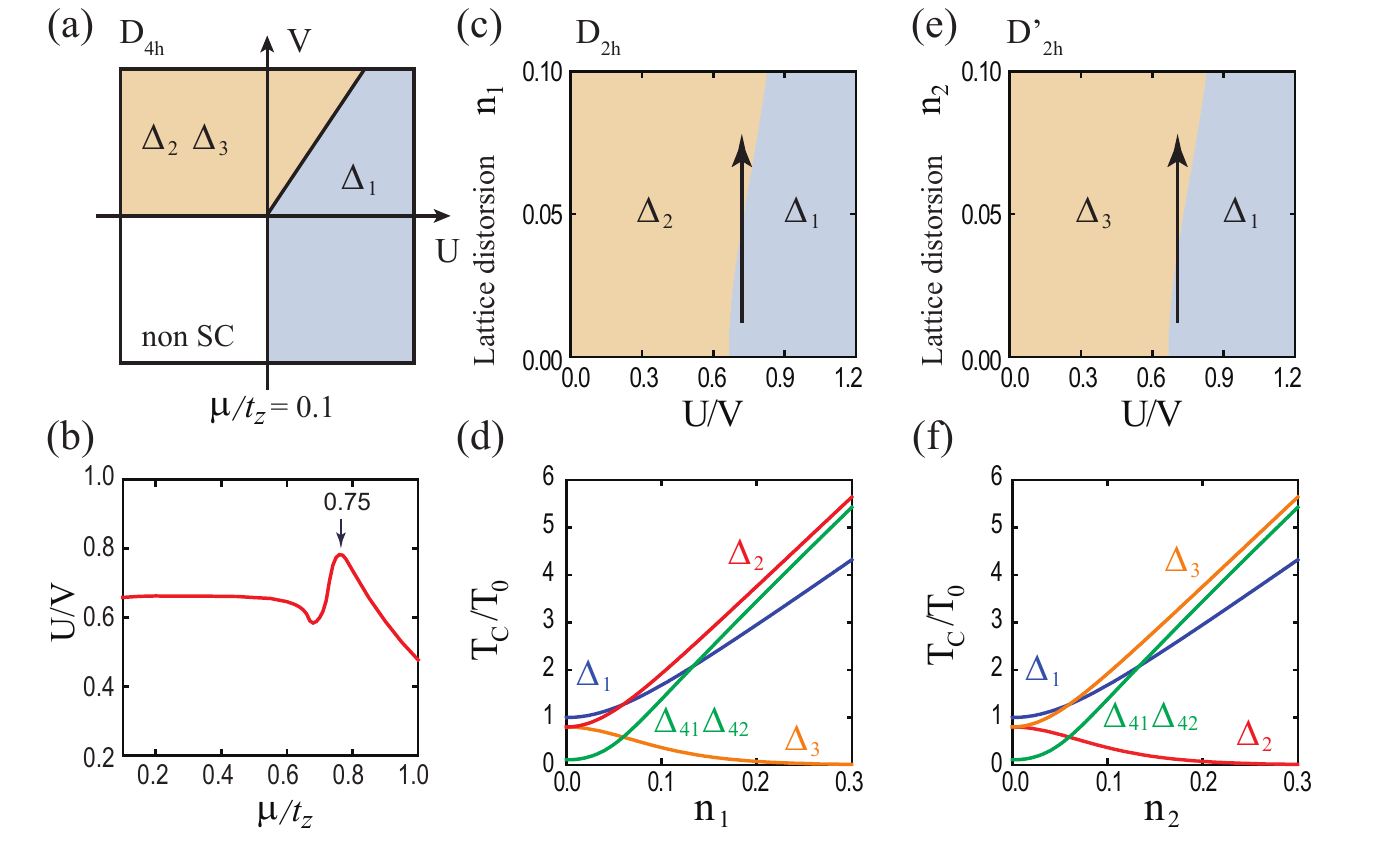} 
\caption{ 
\label{fig5:phase_map_1}
\textbf{Phase diagrams for the tetragonal and orthorhombic crystal systems.}
\textbf{(a)} Superconducting phase diagram in the $U$ and $V$ plane in the absence of lattice distortions. 
In the orange (blue) region, $\Delta_2$ or $\Delta_3$ ($\Delta_1$) phase is dominant.
The slope of the phase boundary is approximately $U/V=2/3$.
The white region indicates a non-superconducting phase.
\textbf{(b)} The numerically calculated critical value of $U/V$ ratio 
as a function of the chemical potential in the absence of lattice distortions.
Since $\mu = 0.75 t_z$ is the band inversion point, there is a local maximum due to Van Hove singularity near $\mu = 0.75 t_z$.
\textbf{(c,e)}
Phase diagrams with respect to (c) $n_1$ and (e) $n_2$ type lattice distortions
when $U=0.045 t_z$.
The corresponding point groups are (c) $D_{2h}$ and (e) $D'_{2h}$.
Each black arrow indicates the possible phase transition from an inversion-even-parity to inversion-odd-parity superconducting phases.
\textbf{(d,f)}
The normalized critical temperature $T_c / T_0$ for various pairing potentials with respect to (d) $n_1$ and (f) $n_2$ type lattice distortions.
In both figures, $U/V=0.75$ and $U=0.045 t_z$}, which corresponds to the black arrows in (c,e).
$T_0$ is the critical temperature of the $\Delta_{1}$ phase in the absence of the lattice distortions.

\end{figure}

First, consider the gap equation in the absence of lattice distortions.
According to the irreducible representation of $D_{4h}$,
$\Delta_{1}$'s, $\Delta_2$, $\Delta_3$ and $\Delta_4$'s belong to $A_{1g}$, $B_{1u}$, $B_{2u}$ and $E_u$ irreducible representations (see Table~\ref{table:pairing_with_lattice_distortion}). 
Then, the gap equations are given by
\begin{align}
&
	\left | \begin{array}{cc}
	U \chi_{1} (T_c) - 1  & U \chi_{1,1'} (T_c) \\
	U \chi_{1,1'} (T_c)  & U \chi_{1'} (T_c) -1
	\end{array} \right |
=  0, ~~~\text{for $\Delta_{1}$ and $\Delta_{1}'$ phases},
\\
&
	V \chi_2 (T_c) - 1 =0,~~~ V \chi_3 (T_c) - 1
= 0, ~~~~~\text{for $\Delta_{2}$ and $\Delta_{3}$ phases},
\\
&
	\left | \begin{array}{cc}
	V \chi_{41} (T_c) - 1  & V \chi_{41,42} (T_c) \\
	V \chi_{41,42} (T_c)  & V \chi_{42} (T_c) -1
	\end{array} \right |
= 0, ~~~~~\text{for $\Delta_{41}$ and $\Delta_{42}$ phases},
\end{align}
where $\chi_{i, j}$ is the generalized superconducting susceptibility for mixed pairings $\Delta_{i}$ and $\Delta_{j}$ by replacing the second $\Delta_{i}$ with  $\Delta_{j}$ in Eq.~(\ref{eq:def_susceptibility}). 
Using the low-energy effective Hamiltonian in Eq.~(\ref{eq:D4h-low-energy-effective-Ham}),
the superconducting susceptibility can be further simplified
and hence one can solve the gap equation analytically.
Using an ellipsoidal coordinate, 
the superconducting susceptibility can be represented as a product of two independent integrals
(see more details in Sec.~S5 in Supplementary Information):
\begin{align}
\chi_i =  \mathcal R (\beta_c) \Omega_i(\mu).
\end{align}
Here, the radial integral part $\mathcal R (\beta_c)$
is given by
\begin{align}
\mathcal R (\beta_c) = 
\int_{-\omega_D}^{\omega_D} 
d E ~  \frac{ \tanh ( \beta_c  E /2) }{ E },
\end{align}
where $E$ is an integration variable and 
$\omega_D$ is the energy cutoff of the pairing potential.
The angular integral part $\Omega_i(\mu)$ 
is given by
\begin{align}
\Omega_i(\mu)
=
\int_{0}^{ \pi} \int_{0}^{ 2 \pi}
\frac{d\theta d\phi }{(2\pi)^3}
\abs{  \frac{ 2  \mu^2  \sin \theta   }{ v^2 v_z } }
f_i( r = \mu,\theta, \phi),
\end{align}
where the form factor $f_i (\mathbf k)$ is represented as a function of $r, \theta$ and $\phi$ in the ellipsoidal coordinates.
After the integration over $\theta$ and $\phi$, the susceptibilities can be obtained as follows:
\begin{align}
\chi_{1} = 4 \pi C_0 \mathcal R(T_c),
~
\chi_{1'}
=  \frac{4 \pi}{3}  C_0  \mathcal R(T_c),
~
\chi_{2}
=  \chi_{3} = \frac{8 \pi}{3} C_0 \mathcal R(T_c),
~
\chi_{41}
=  \chi_{42} =\frac{4 \pi}{3} C_0 \mathcal R(T_c),
~
\chi_{1,1'} = \chi_{41,42}=0,
\end{align}
where
$ C_0 = \frac{2}{(2\pi)^3}\frac{\mu ^2}{v^2 v_z } $.
Then, the linearized gap equations are given by
\begin{align}
& \chi_{1} (T_c) = \chi_{1'}(T_c) = 1/U,\\
& \chi_2 (T_c) = \chi_3 (T_c) = \chi_{41} (T_c) = \chi_{42} (T_c)  = 1/V.
\end{align}
If we denote the critical temperature $T_c^{(i)}$ for a pairing potential $\Delta_i$, then
the gap equations are given by
\begin{align}
& \mathcal R(T^{(1)}_c)  = \frac{1}{3} \mathcal R(T^{(1')}_c)  = \frac{1}{4 \pi  U C_0},\\
& \mathcal R(T^{(2)}_c)  = \mathcal R(T^{(3)}_c)  =  \frac{1}{2} \mathcal R(T^{(41)}_c)  = \frac{1}{2} \mathcal R(T^{(42)}_c) = \frac{3}{ 8\pi  V C_0}.
\end{align}
Because $\mathcal R(x) $ is a monotonically decreasing function with respect to $x$,
$T_{c}^{(1)} > T_{c}^{(1')} $ and
$T_{c}^{(2)} = T_{c}^{(3)} >  T_{c}^{(41)} = T_{c}^{(42)}$.
Thus, the highest $T_c$ is determined among $T_{c}^{(1)}, T_{c}^{(2)}$ and $T_{c}^{(3)}$.
Because the critical temperatures are same at the phase boundary,
the phase boundary in Fig.~\ref{fig5:phase_map_1}(a) is determined by the equation
$\mathcal R(T^{(1)}_c) = \mathcal R(T^{(2)}_c) = \mathcal R(T^{(3)}_c)$,
which gives the critical value of $U/V = 2/3$.

When the chemical doping is low, the superconducting phase diagram for undistorted Dirac semimetal is shown in Fig.~\ref{fig5:phase_map_1}(a).
When the intra-orbital interaction $U$ is strong, the conventional $s$-wave  superconductivity with pairing potential $\Delta_{1}$ is the dominant phase.
However, with the increasing inter-orbital interaction $V$, the unconventional superconducting phase with inter-orbital pairing potential $\Delta_{2}$ or $\Delta_{3}$ can emerge.
Figure~\ref{fig5:phase_map_1}(b) shows the numerically obtained critical value of $U/V$ ratio using the lattice Hamiltonian.
Thus, by controlling the $U/V$ ratio, both conventional and unconventional superconductivity can emerge for for the large range of chemical doping.
The calculated value of $U/V$ ratio is similar with $2/3$ using the low-energy effective Hamiltonian, which means that $\Delta_{2}$ or $\Delta_{3}$ phase can emerge for the large range of chemical doping.

Next, consider the effect of $n_1$ and $n_2$ types of lattice distortions on the superconducting temperatures and the phase diagrams.
When $n_1$ type lattice distortion is turned on, the point group becomes $D_{2h}$.
In this case, only $\Delta_{1}$ and $\Delta_{1}'$ belong to the same $A_{g}$ class, and the others are belong to different classes (see Table~\ref{table:pairing_with_lattice_distortion}).
So the linearized gap equation is given by
\begin{align}
&
	\left | \begin{array}{cc}
	U \chi_{1} (T_c) - 1  & U \chi_{1,1'} (T_c) \\
	U \chi_{1,1'} (T_c)  & U \chi_{1'} (T_c) -1
	\end{array} \right |
=  0, ~~~\text{for $\Delta_{1}$ and $\Delta_{1}'$ phases},
\\
&
V \chi_2 (T_c) =
V \chi_3 (T_c) =
V \chi_{41} (T_c) =
V \chi_{42} (T_c) = 1,~~~\text{for $\Delta_{2}$, $\Delta_{3}$, $\Delta_{41}$, and $\Delta_{42}$phases}.
\end{align}
Similar to $D_{4h}$ case, the susceptibility can be analytically calculated when the chemical doping level is small.
The relevant gap equations that determine the phase map are given by
\begin{align}
\mathcal R(T^{(1)}_c)  = \frac{1}{4 \pi U C_0},~~~~~~
\mathcal R(T^{(2)}_c)  = \frac{1}{\frac{4\pi}{3} (2 + \frac{n_1^2 \sin ^2 k_0}{\mu^2}) V C_0}.
\end{align}
Thus, the phase boundary is given by
\begin{align}
\frac{U}{V} = \frac{2 + (n_1^2 \sin^2 k_0)/{\mu^2} }{3}.
\end{align}
Similarly, the other cases can be calculated. See the details in Supplementary Information.

Figure~\ref{fig5:phase_map_1} shows the numerically calculated phase maps under the $n_1$ and $n_2$ types of lattice distortions using the low-energy effective Hamiltonian.
The phase diagrams are plotted in the plane of the $U/V$ ratio versus strength of $n_1$ or $n_2$ type lattice distortion. In each diagram, the dominant phases are conventional  spin-singlet $\Delta_{1}$ phase and unconventional spin-triplet $\Delta_2$ or $\Delta_3$ phase depending on the parameters.
When $U/V$ is small (large) enough, $\Delta_2$ or $\Delta_3$ ($\Delta_{1}$) phase emerges.
Remarkably, the unconventional superconductivity can emerge with increasing lattice distortions.
As an example, near the phase boundary of $U/V \approx 0.7$, there is a phase transition between conventional superconducting $\Delta_{1}$ and unconventional superconducting $\Delta_{2}$ phases when $n_1$ increases [see the black arrow in Fig.~\ref{fig5:phase_map_1}(c)].
To see this phase transition more clearly, we plot the normalized superconducting critical temperatures along the black arrow [Fig.~\ref{fig5:phase_map_1}(d)].
When $n_1=0$, the $\Delta_{1}$ phase is dominant.
With increasing $n_1$, the superconducting critical temperatures for $\Delta_{2}$ are increasing, which leads to the $\Delta_{2}$ superconducting phase under enough lattice distortion.
Note that $T_c$'s for $\Delta_{1}$, $\Delta_{2}$, $\Delta_{41}$, and $\Delta_{42}$ increase while $T_c$ for $\Delta_{3}$ decreases with the increasing $n_1$ [Fig.~\ref{fig5:phase_map_1}(d)].
This can be explained by the expectation values of the Cooper pairings and spin-orbital texture at the Fermi surface, which will be discussed later.
Because $n_1$ and $n_2$ type lattice distortions are related with $\pi/4$ rotation, similar features are observed except for the exchange of $\Delta_2$ and $\Delta_3$ phases [Fig.~\ref{fig5:phase_map_1}(e,f)].

\begin{figure}[!t]
\centering
\includegraphics[width=120mm]{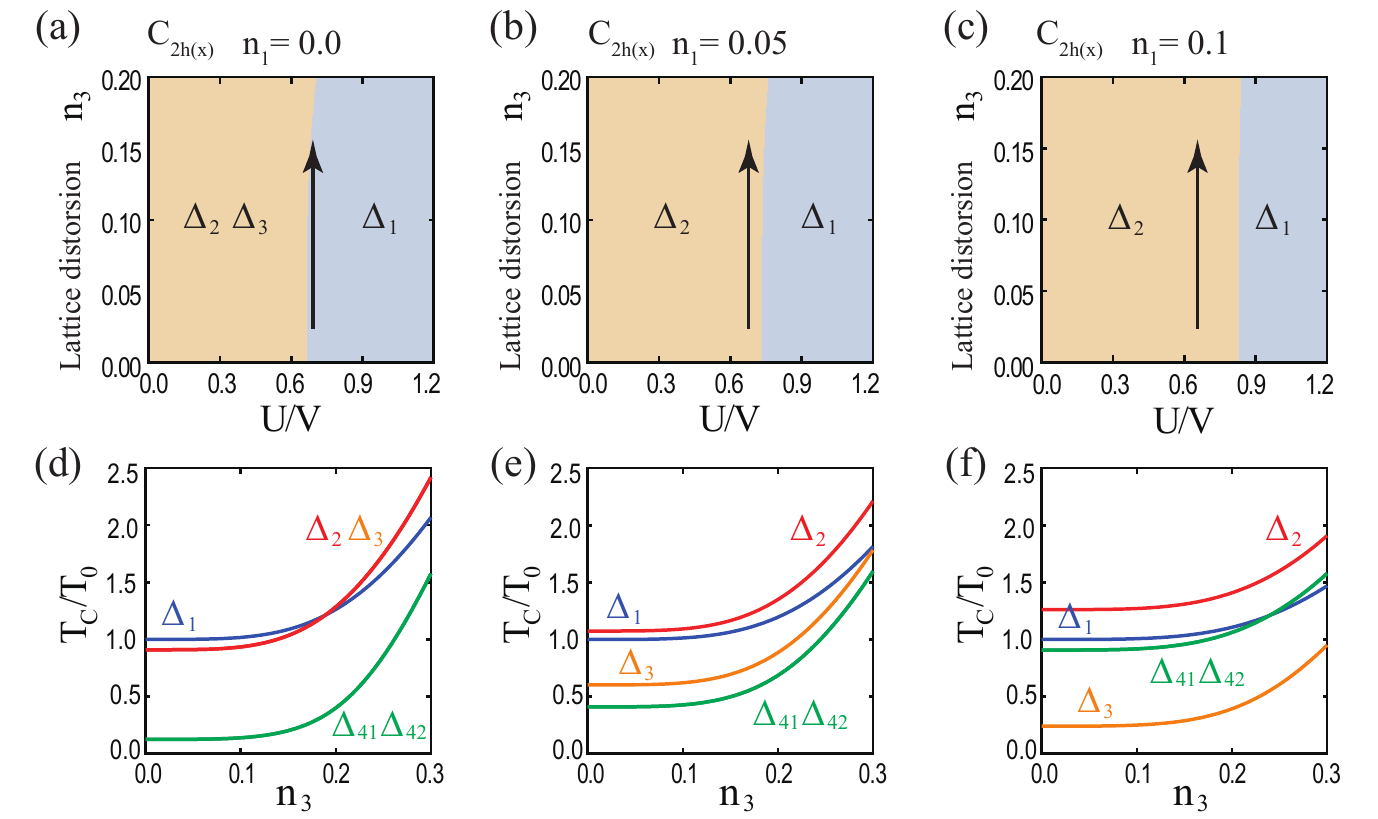} 
\caption{ 
\label{fig6:phase_map_2}
\textbf{Phase diagrams for the monoclinic crystal system.}
\textbf{(a-c)}
Phase diagrams with respect to $U/V$ ratio and $n_3$ type lattice distortions
for (a) $n_1=0.0$, (b) $n_1=0.05$, and (c) $n_1=0.1$.
\textbf{(d-f)}
The normalized critical temperature $T_c / T_0$ along the black arrows in (a-c).
Here, $U/V=0.7$ and $T_0$ is the critical temperature of the $\Delta_{1}$ phase in the absence of the lattice distortions.
In (d), the red and orange lines for $\Delta_{2}$ and $\Delta_{3}$ overlap.
} 
\end{figure}

For the $n_3$ type lattice distortion, similar features can be observed in Fig.~\ref{fig6:phase_map_2}.
Under the $n_3$ type lattice distortion, $n_1$ type lattice distortions also can be involved as discussed before.
Thus, we plot three representative phase diagrams for $n_1=0.0$, $0.05$, and $0.1$.
Surprisingly, when $n_1=0$, $\Delta_{2}$ and $\Delta_{3}$ phases are degenerate, and they are dominant unconventional phases as shown in Fig.~\ref{fig6:phase_map_2}(a,d). 
With increasing $n_1$, the region of the unconventional phase $\Delta_{2}$ increases [Fig.~\ref{fig6:phase_map_2}(a-c)]
and the degenerate $\Delta_{2}$ and $\Delta_{3}$ phases become distinguishable.

Under $n_1$, $n_2$, and $n_3$ lattice distortions, the $T_c$'s of $\Delta_{2}$, $\Delta_{3}$, $\Delta_{41}$, and $\Delta_{42}$ increases much more than that of $\Delta_{1}$ [Figs.~\ref{fig5:phase_map_1}(d,f), \ref{fig6:phase_map_2}(d-f)], and hence the unconventional superconducting phases emerge. The mechanism of this will be discussed below.

\subsection*{Mechanism for $T_c$ enhancement of unconventional superconductivity}
The $T_c$ enhancement of unconventional
superconductivity under lattice distortions can be understood by the enhancement of DOS at Fermi surface and the enhancement of the expectation values of unconventional pairings at Fermi surfaces due to the unique spin-orbital texture.

First, we consider the increment of DOS at the Fermi surface.
Under the lattice distortions, the DOS's at the Fermi surface increase as shown in Eqs.~(\ref{eq:DSM_DOS_01}) and (\ref{eq:DSM_DOS_02}).
Then, the superconducting critical temperature increases under lattice distortions because $T_c \propto e^{- \frac{1}{g N(0)} }$.
Here, $g$ is the strength of the pairing potential in the standard BCS theory and $N(0)$ is the DOS at Fermi surface.
Due to this enhancement of DOS, most of the superconducting temperatures  increase under the lattice distortions [see Figs.~\ref{fig5:phase_map_1}(d,f) and \ref{fig6:phase_map_2}(d,e,f)].
However, some unconventional superconducting temperatures decrease while some unconventional superconducting temperatures increase under lattice distortions. To understand this, we investigate the pairing expectation values for each superconducting pairing potentials.

\begin{figure}[!t]
\centering
\includegraphics[width=160mm]{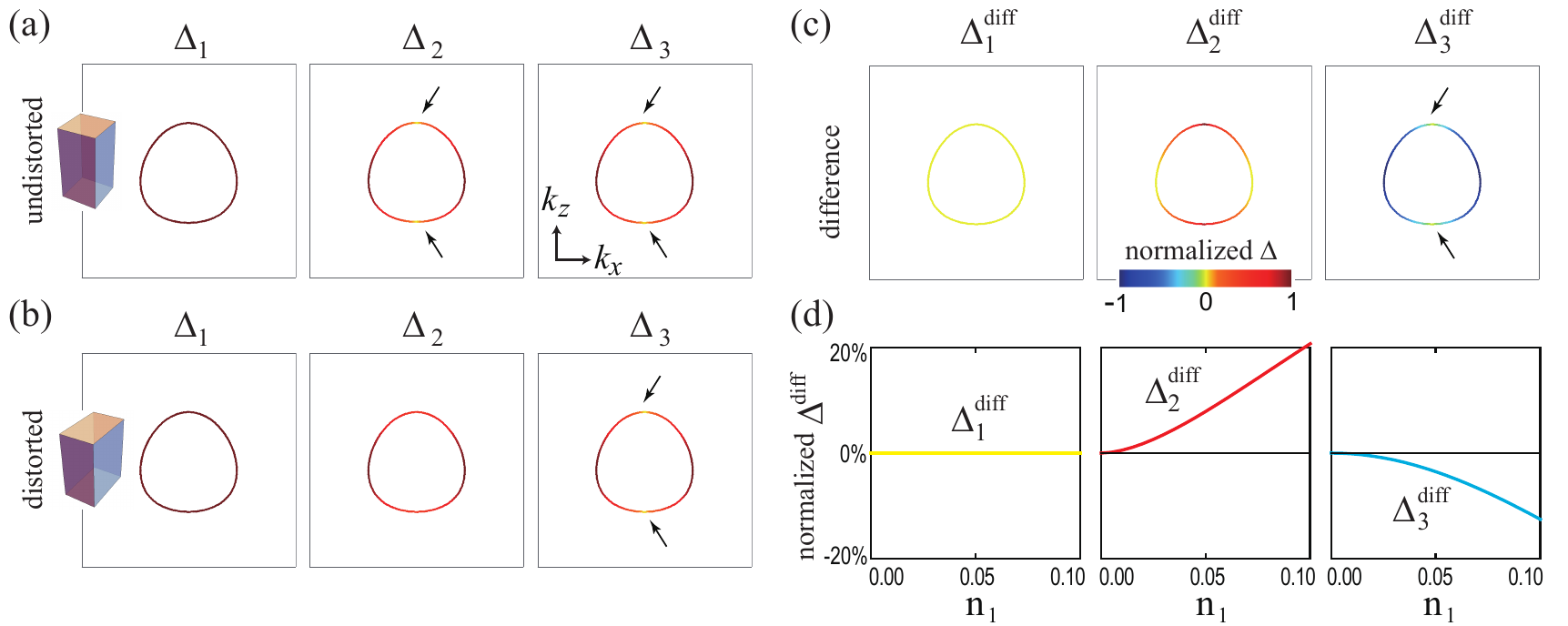} 
\caption{ 
\label{fig7:expectation_value}
\textbf{Expectation values of pairing potentials at the upper Fermi surface under the $n_1$ type lattice distortion.}
\textbf{(a,b)}
The normalized expectation values $\Braket{\Delta_i}$ of $\Delta_{1}$, $\Delta_{2}$, and $\Delta_{3}$ (a) without and (b) with the $n_1$ type lattice distortion are plotted at the upper Fermi surface of DSM
in the $k_x$-$k_z$ plane.
\textbf{(c)} 
The differences $\Delta_i^{\text{diff}} \equiv \Braket{\Delta_i}_{n_1 \neq 0} - \Braket{\Delta_i}_{n_1=0}$  are plotted.
In (a-c), the black arrows indicate the points having zero expectation values.
\textbf{(d)} 
The normalized integrated expectation values of each pairing potentials, $\int_{\text{FS}} d^2 k  \Delta_i^{\text{diff}} / \int_{\text{FS}} d^2 k  \Braket{ \Delta_i}_{n_1=0} $,  are plotted with respect to $n_1$.
Note that the upper Fermi surfaces encloses the Dirac point $( 0, 0,  k_0)$ as shown in Fig.~\ref{fig1:lattice_distortions}(k-o).
} 
\end{figure}

As a representative example, we calculate the normalized expectation values for the $\Delta_{1}$, $\Delta_2$, and $\Delta_3$ pairings at the Fermi surface with and without the $n_1$ type lattice distortion [Figure~\ref{fig7:expectation_value}(a,b)].
For a clear comparison, the differences $\Delta_i^{\text{diff}} \equiv \Braket{\Delta_i}_{n_1 \neq 0} - \Braket{\Delta_i}_{n_1=0}$ are calculated [Fig.~\ref{fig7:expectation_value}(c)].
Without lattice distortions, $\Braket{\Delta_{1}}$ is uniform while $\Braket{\Delta_{2}}$ and $\Braket{\Delta_{3}}$ show zeros on the $k_z$ axis.
With the $n_1$ type lattice distortion,
$\Braket{\Delta_{2}}$ increases while $\Braket{\Delta_{3}}$ decreases,
which leads to $\Delta_{2}^{\text{diff}} > 0 $ and $\Delta_{3}^{\text{diff}} <0 $ [Fig.~\ref{fig7:expectation_value}(c,d)].
On the other hand, $\Delta_{1}^{\text{diff}}= 0$.
These behaviors of the expectation values of $\Braket{\Delta_{i}}$ explains that the tendency of $T_c$ under lattice distortions.
$T_c$ of $\Delta_{2}$ phase increase greater than that of $\Delta_{1}$ phase while $T_c$ of $\Delta_{3}$ phase decreases
under $n_1$ type lattice distortion [Fig.~\ref{fig5:phase_map_1}(a)].
Similarly, the effect of the other types of lattice distortions on $T_c$ can be understood by the expectation value change of the pairing potentials.

Microscopically, we can understand the emergence of unconventional superconducting phases under lattice distortions as a result of the enhancement of inter-orbital pairing at the Fermi surface.
Even though our argument can be applied to all distortions, we discuss the effect of $n_1$ type lattice distortion for convenience.
We consider two Fermi surfaces encapsulating Dirac points $( 0, 0, \pm k_0)$ which are related by time-reversal and inversion.
On the upper Fermi surface near the Dirac point $( 0, 0, + k_0)$, 
the Dirac Hamiltonian in Eq.(\ref{eq:Ham_Dirac_n1_n2}) in the $k_x$-$k_z$ plane is given by  
\begin{align}  \label{eq:Ham_xz_plane}
H_{\text{Dirac}} =  ( n_1 \sin k_0 s_x +  v k_x s_z ) \sigma_x + v_z (k_z-k_0) \sigma_z.
\end{align}
The spin and orbital parts can be diagonalized separately
and the total wavefunction can be represented by the product of spin and orbital wavefunctions\cite{hashimoto2016superconductivity}:
\begin{align} 
\Ket { \Psi } = \Ket { \phi }_{\text{orbital}}  \otimes \Ket { \psi }_{\text{spin}}.
\end{align}
Let us diagonalize the spin part.
The spin part of Hamiltonian is given by
\begin{align} 
H_{\text{spin}} =  ( n_1 \sin k_0 s_x +  v k_x s_z ) = \mathbf{h} \cdot \mathbf {s},
\end{align}
where  $ \mathbf{h} = (n_1 \sin k_0 , 0,  v k_x)$.
Since this Hamiltonian is a product of momentum and spin operators,
the spin wavefunction can be represented in the helicity basis $ \Ket{ \lambda}_{\text{spin}}$ with $\lambda=\pm1$:
\begin{align} 
H_{\text{spin}} \Ket{ \lambda }_{\text{spin}} & =   \lambda \abs{ \mathbf{h} } \Ket{ \lambda }_{\text{spin}}.
\end{align}

Next, we diagonalize the remaining orbital part.
Depending on the spin helicity $\lambda$, the Hamiltonian in Eq.~(\ref{eq:Ham_xz_plane}) can be written as follows:
\begin{align} 
  H_{\text{orbital},~\lambda}
  =  \lambda\abs{\mathbf h} \sigma_x + v_z (k_z-k_0) \sigma_z
  =  \mathbf{ d}_{\lambda}\cdot \boldsymbol{\sigma},
\end{align}
where $\mathbf{d}_{\lambda} = ( \lambda \abs{\mathbf h}, 0, d_z)$ and $d_z = v_z (k_z-k_0)$.
The orbital wavefunction can be represented by the pseudo-spin along $ \mathbf {\hat d_{\lambda}} $.
For each spin helicity $\lambda$, there are two orbital wavefunctions $\Ket{ \kappa \mathbf {\hat d_{\lambda}}}_{\text{orbital}} $ with $\kappa=\pm1$ that satisfy the following equations:
\begin{equation} 
H_{\text{orbital},~\lambda}  \Ket{ \kappa \mathbf {\hat d_{\lambda}} }_{\text{orbital}} 
=  \kappa d  \Ket{ \kappa \mathbf {\hat d_{\lambda}} }_{\text{orbital}},
\end{equation}
where $d =  \sqrt{ \abs{\mathbf h} ^2 + d_z^2 } $.
When the chemical potential is positive, two degenerate wavefunctions located in conduction bands participate in the superconducting pairing.
These wavefunctions are given by
\begin{align} \label{eq:wavefunction_PT_pair}
\Ket{\Psi_1} = \Ket{\mathbf{\hat d} _{+}}_{\text{orbital}} \otimes  \Ket{ +  }_{\text{spin}},~~~~~~~~
\Ket{\Psi_2} = \Ket{\mathbf{\hat d} _{-}}_{\text{orbital}} \otimes  \Ket{ -  }_{\text{spin}},
\end{align}
which form a Kramer's pair due to the $PT$ symmetry regardless of lattice distortions:
$PT$ operation conserves the momentum while it flips helicity and the $x$-component of the orbital because $T= i s_y \hat K$ and  $P=-\sigma_z$.

\begin{figure}[!t]
\centering
\includegraphics[width=160mm]{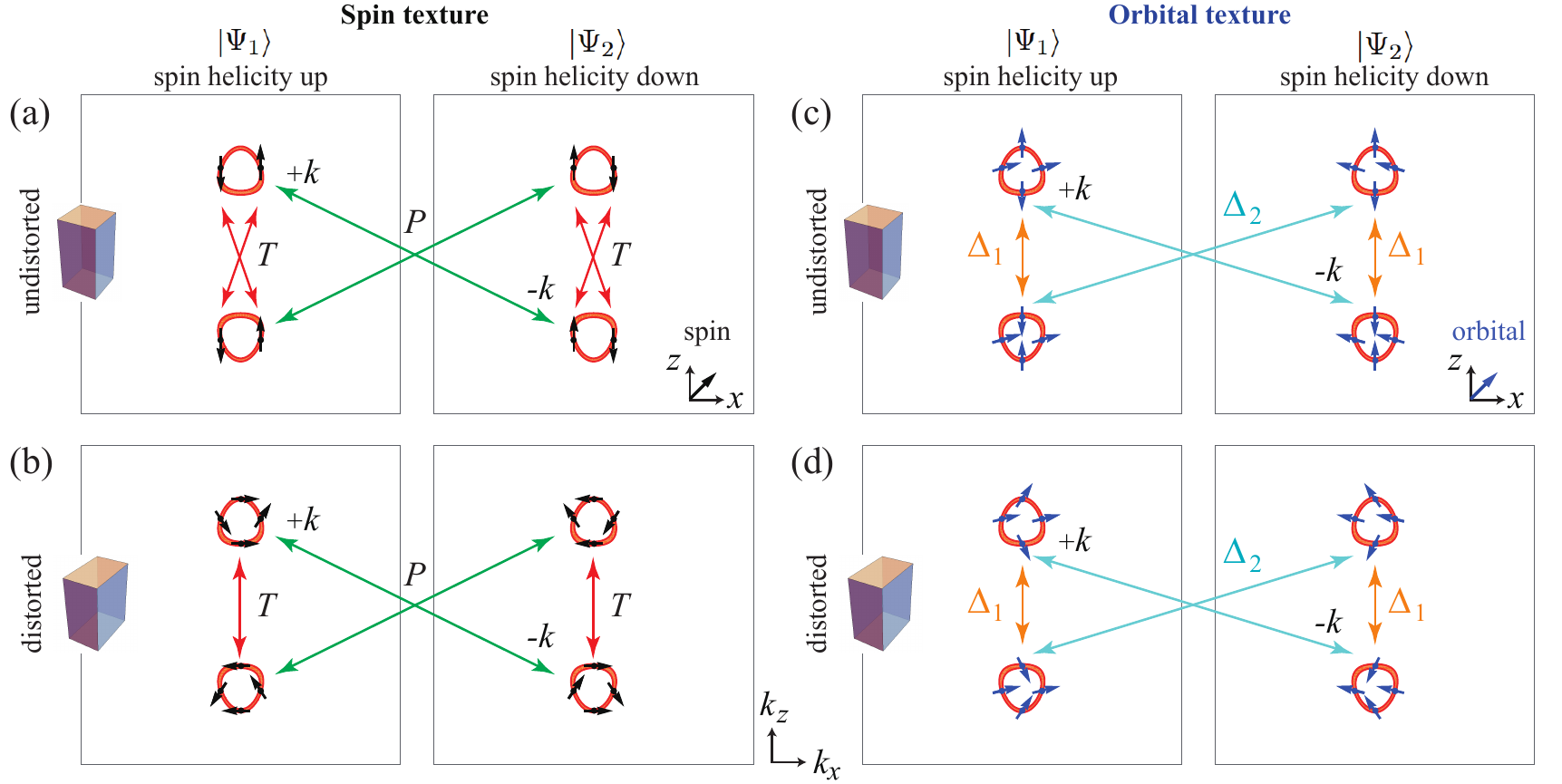} 
\caption{ 
\label{fig8:orbital_texture}
\textbf{Spin and orbital textures without and with lattice distortion.}
\textbf{(a,b) [(c,d)]} Numerically calculated spin (orbital) textures at two Fermi surface surfaces.
The $n_1$ type lattice distortion is absent in (a,c) and present in (b,d).
In (a,b) [(c,d)], the spin (orbital) textures are represented by the small black (blue) arrows.
In (a-d), the textures in left and right panels correspond to the spin helicity up and down wavefunctions, respectively. In (a,b), the red and green arrows indicate time-reversal and inversion pairs, respectively. In (c,d), the orange and blue arrows indicate the possible Cooper pairing between two electrons with opposite momenta.
Note that the orbital pseudo-spin vectors connected by orange arrows are parallel regardless of the lattice distortion.
On the other hand, the orbital pseudo-spin vectors connected by cyan arrows are parallel in (c) while non-parallel in (d).
} 
\end{figure}

Since we have obtained the spin and orbital texture in one Fermi surface, we can obtain the spin and orbital texture of the other Fermi surface by applying either time-reversal or inversion operator.
Let $\Psi(\mathbf k)$ be a wavefunction on the Fermi surface.
Because there is no $\sigma_y$ in the Hamiltonian Eq.~(\ref{eq:Ham_xz_plane}), the time-reversal partner $ T \Psi (\mathbf k) $ has the same orbital direction and the opposite spin direction regardless of lattice distortions comparing with $\Psi(\mathbf k)$.
On the other hand, since $P=-\sigma_z$, the inversion partner $ P \Psi (\mathbf k) $ has the opposite $d_x$  while keeping $d_z$ and spin direction comparing with $\Psi(\mathbf k)$.
Figure \ref{fig8:orbital_texture} shows the numerically calculated spin and orbital textures using the lattice model.
The $P$ and $T$ symmetry operators connects spin and orbital wavefunctions in Fig.~\ref{fig8:orbital_texture}.
The red and green arrows indicate time-reversal and inversion pairs, respectively.

Using these spin and orbital textures, let us investigate how the lattice distortions promote the unconventional pairings.
The conventional $\Delta_{1}$ pairing is not affected by the lattice distortion.
The expectation value of $\Delta_{1}$ is constant over the entire Fermi surface regardless of lattice distortion as shown in Fig.~\ref{fig7:expectation_value}(a,b).
Because $\Delta_{1}= c^{\dag}_{1\uparrow}c^{\dag}_{1\downarrow}+c^{\dag}_{2\uparrow}c^{\dag}_{2\downarrow} + H.c. $
connects two wavefunctions that are related by time-reversal, the expectation value of $\Delta_{1}$ is constant due to TRS.
In other words, because $\Delta_{1}$ is represented by the identity matrix $1_{4 \times 4}$, the expectation value of the $\Delta_{1}$ over the Fermi surface is constant even under the lattice distortions.

On the other hand, $n_1$ type lattice distortion can increase the expectation values of the inter-orbital pairing $\Delta_{2}$.
For  example, let us consider two wavefunctions located at the south pole of the upper Fermi surface ($k_z = + k_s$ with $k_s<k_0$) and the north pole of the lower Fermi surface ($k_z=-k_s$). Two wavefunctions are indicated by the orange and cyan arrows in Fig.~\ref{fig8:orbital_texture}(c,d).
At $k_z = \pm k_s$, the Dirac Hamiltonian in Eq.(\ref{eq:Ham_Dirac_n1_n2}) is given by
\begin{align}  
H_{\text{Dirac}} ^{(\pm)} =  \pm n_1 \sin k_0 s_x  \sigma_x  -  v_z (k_0 - k_s) \sigma_z,
\end{align}
where $H_{\text{Dirac}} ^{(+)}$ and $H_{\text{Dirac}} ^{(-)}$ correspond $k_z=k_s$ and $k_z=-k_s$, respectively.
When $n_1=0$, wave functions on the conduction bands at $k_z=\pm k_s$  are given by
\begin{align} 
& \Ket { 2 }_{\text{orbital}}  \otimes \Ket { \uparrow_{x} }_{\text{spin}},
~~~
\Ket { 2 }_{\text{orbital}}  \otimes \Ket { \downarrow_{x} }_{\text{spin}},
~~~ \text{for} ~ k_z= + k_s, \label{eq:78:wavefuntion-orbital-texture-1}
\\
& \Ket { 2 }_{\text{orbital}}  \otimes \Ket { \uparrow_{x} }_{\text{spin}},
~~~
\Ket { 2 }_{\text{orbital}}  \otimes \Ket { \downarrow_{x} }_{\text{spin}},
~~~ \text{for} ~ k_z=-k_s, \label{eq:79:wavefuntion-orbital-texture-2}
\end{align}
where $\Ket { 1 }_{\text{orbital}}$ and $ \Ket {2 }_{\text{orbital}}$ indicate the orbital basis for $\sigma$ matrix as defined before.
$\Ket { \uparrow_{x} }_{\text{spin}} $ and $\Ket { \downarrow_{x} }_{\text{spin}}$ indicate the spin up and down along $x$-direction.
Thus, the expectation value of inter-orbital pairing is zero for these wavefunctions because the orbital states of the wavefunctions in Eqs.~(\ref{eq:78:wavefuntion-orbital-texture-1}) and, (\ref{eq:79:wavefuntion-orbital-texture-2}) are same.
On the other hand, when $n_1 \neq 0$, the $x$-component of the orbital pseudo-spin is generated [indicated in the large cyan arrows in Fig.~\ref{fig8:orbital_texture}(d)]. 
The wave functions at $k_z=\pm k_s$ are given by
\begin{align} 
&\Ket{\hat{\mathbf{d}}_{+}}_{\text{orbital}}\otimes\Ket{\uparrow_{x}}_{\text{spin}},~~~
\Ket{\hat{\mathbf{d}}_-}_{\text{orbital}}\otimes\Ket{\downarrow_{x}}_{\text{spin}},~~~ \text{for} ~ k_z=+k_s, \\
&\Ket{\hat{\mathbf{d}}_-}_{\text{orbital}}\otimes\Ket{\uparrow_{x}}_{\text{spin}},~~~
\Ket{\hat{\mathbf{d}}_+}_{\text{orbital}}\otimes\Ket{\downarrow_{x}}_{\text{spin}},~~~ \text{for} ~ k_z=-k_s,
\end{align}
where $ \mathbf{d}_{\pm} = ( \pm n_1 \sin k_0 , 0,  - v_z(k_0 - k_s))$.
Therefore, under the lattice distortion, the expectation value of the inter-orbital pairing is allowed and $\Delta_{2}$ pairing is enhanced.
This mechanism for the enhancement of unconventional pairings can be applied to the other cases.
In summary, the emergence of unconventional superconductivity under lattice distortion can be understood due to the enhancement of inter-orbital pairings and DOS at Fermi surfaces.

\subsection*{Topological superconductivity of doped Dirac semimetal}

\begin{table}[t]
\centering
\begin{tabular}{ c c l c c c c c c}
\hline 
\hline
Type & Gap & Class & Topological invariant & Classification &  $\Delta$'s\\
\hline 
\hline
Line nodal SC & LN & DIII + P$_\text{even}$ & $W$ & $2 \mathbb Z $ &  $\Delta_{1}, \Delta_{1}' $
\\
Topological mirror SC & FG & DIII + P$_\text{odd} + $ M$_{\text{odd}}$ & $C_M$ & $ 2 \mathbb Z $ & $\Delta_{2}, \Delta_{3} $
\\
Point nodal SC & PN & DIII + P$_\text{odd} + $ M$_{\text{even}}$ & $W_{M}$ & $2 \mathbb Z $ & $\Delta_{2},\Delta_{3},\Delta_{41},\Delta_{42} $
\\
\hline
\hline
\end{tabular}
\caption{
\label{table:various_superconductivity}
Possible topological superconductivity in doped DSM under lattice distortions.
SC, FG, LN, and PN denote superconductor, full gap, line node, and point node, respectively.
P$_\text{odd}$ and P$_\text{even}$ represent the inversion-odd and inversion-even parity superconductors.
M$_\text{odd}$ and M$_\text{even}$ represent the mirror-odd and mirror-even parity superconductors.
$C_{M}$ is the mirror Chern number.
$W $ is chiral winding number defined by Eq.~(\ref{eq:chiral_winding_number_def}).
$W_{M} $ is the mirror chiral winding number defined by Eq.~(\ref{eq:mirror_chiral_winding_number}).
Here, the $2\mathbb{Z}$ indicates the even number of the corresponding surface Andreev bound state (SABS).
} 
\end{table}

As summarized in Table~\ref{table:various_superconductivity}, we characterize possible superconducting states in doped Dirac semimetal by the gap structures, topological winding numbers, and surface spectra.

First, the conventional superconducting phase having $\Delta_{1}$ pairing potential can emerge. Because $T_c$ of the $\Delta_{1}$ phase increases under lattice distortions as shown in Figs.~\ref{fig5:phase_map_1} and \ref{fig6:phase_map_2}, conventional fully-gapped $s$-wave superconductivity can emerge. 

Second, we consider the inversion-odd-parity superconductor.
The BdG Hamiltonian in Eq.~(\ref{eq:BdG_Hamiltonian}) are included in the DIII class according to 10-fold Altland-Zirnbauer classes~\cite{schnyder2008classification, chiu2016classification}
because  $T^2 = -1, C^2 = +1$, and $\Gamma^2 = +1$.
With the additional inversion symmetry,
the DIII class superconductor can be an inversion-odd-parity topological superconductor\cite{fu2010odd} classified by $Z_2$ invariants $ (-1) ^ { w_{ \text{DIII} } }$, where
\begin{align} \label{eq:DIII_winding_number}
w_{\text{DIII}} = - \int \frac{d^3 k}{48 \pi^3} \epsilon_{\mu \nu \rho} 
\text{Tr}[ \Gamma (Q \partial_{\mu} Q) (Q \partial_{\nu} Q) (Q \partial_{\rho} Q) ].
\end{align}
Here, $\Gamma$ is the chiral operator, and $Q$ is the so-called $Q$-matrix\cite{schnyder2008classification, chiu2016classification} (or projection matrix).
The sufficient condition for realizing the inversion-odd-parity topological superconductor is 
that it has an inversion-odd-parity pairing with a full gap and its Fermi surface encloses an odd number of time-reversal-invariant momenta.
In the absence of lattice distortions, the inversion-odd-parity pairings, $\Delta_{2}$, $\Delta_{3}$, $\Delta_{41}$, and $\Delta_{42}$, are not fully gapped [Fig.~\ref{fig2:full_nodal_structure}(a)] and cannot be such a topological superconductor.
However, under the lattice distortions, 
these inversion-odd-parity phases can be fully gapped, and the sufficient condition above can be satisfied for the large chemical potential ($\mu > M_0$) because the Fermi surface can enclose only $(0,0,0)$ in BZ.
However, when the chemical potential is large with a lattice distortion, the band structure near the Fermi energy is far from that of DSM.
Because we are discussing the Dirac physics, we do not consider such a superconducting phase in this work.

Third, topological mirror superconducting phases\cite{zhang2013topological,kobayashi2015topological} can exist under lattice distortions.
Topological DSM has a nontrivial mirror Chern number defined in the $M_{xy}$ plane and the corresponding surface states on the mirror-symmetric boundary\cite{yang2014classification, kobayashi2015topological}.
Similarly, topological mirror superconductivity for $\Delta_{2}$ and $\Delta_{3}$ phases can exist under lattice distortions.
Under the $n_1$ ($n_2$) type lattice distortion, $\Delta_{2}$ ($\Delta_{3}$) phase is fully gapped, the $\Delta_{2}$ ($\Delta_{3}$) pairing potential is mirror-odd under the $M_{xy}$ symmetry, and the  mirror Chern number defined in $M_{xy}$ plane is nontrivial ($C_{M} = \pm 2$),
which leads to topological mirror superconductivity with a topologically-protected Majorana states on the mirror symmetric boundary.
For example, see the gapless surface spectra of $\Delta_{2}$ and $\Delta_{3}$ phases in Region III in Fig.~\ref{fig4:surface_spectrum}(a,b,e,f).
Due to the TRS and IS, this topological mirror superconductor is classified as $2\mathbb{Z}$. 

Fourth, topological line nodal superconducting phases can exist under lattice distortions.
As discussed in Fig~\ref{fig3:typical_winding_number}(b,c), the inversion-even-parity $\Delta_{1}'$ pairing allows a topologically-protected nodal lines protected by the chiral winding number in Eq.~(\ref{eq:chiral_winding_number_def}).
According to this chiral winding number, in general, the topological line nodal superconductor in doped topological DSM is classified as $2\mathbb{Z}$.
The reason is as follows.
Since there are $PT$ and $PC$, the nodal points are fourfold degenerate,
which means that there are even number of winding source at the same points.
Therefore, our generic model has a topological winding number of even integers.
Note that the topological class of a line node in 3D DIII superconductor using Clifford algebra\cite{kobayashi2014topological} is $2\mathbb{Z}$, which is consistent with our result. However, there is no surface state because $\Delta_{1}'$ phase has two nodal lines having opposite chiral winding numbers [Fig.~\ref{fig3:typical_winding_number}(a)].

Fourth, topological point nodal superconducting phases can exist under lattice distortions.
For an inversion-odd-parity and mirror-even-parity pairing potential, we have a topological point nodal superconductor of which nodal points are protected by the mirror chiral winding number in Eq.~(\ref{eq:mirror_chiral_winding_number}).
Because the chiral winding number is zero for inversion-odd-parity superconductor ($W = W_{\lambda=i} + W_{\lambda=-i} = 0$),
the mirror chiral winding number is given by $ W_{M} = W_{\lambda=i} -  W_{\lambda=-i} = 2 W_{\lambda=i}$.
From this mirror chiral winding number, this topological point nodal superconductor is classified as $2\mathbb{Z}$. 
Note that the classification of a point node using Clifford algebra\cite{kobayashi2014topological,chiu2016classification} is $M\mathbb Z$ considering one mirror sector, which is consistent with our results.

\section*{\label{sec:discussion}Discussion}
Now, we compare our results with experimental works in doped DSM of Au$_2$Pb\cite{schoop_dirac_2015,chen_temperature-pressure_2016,xing_superconductivity_2016,yu2017fully,wu2019ground} and Cd$_3$As$_2$\cite{aggarwal_unconventional_2016,he_pressure-induced_2016,wang2016observation}.
Au$_2$Pb shows superconductivity at $T_c \approx 1.2$K with $D_{2h}$ symmetry at the ambient pressure\cite{schoop_dirac_2015,xing_superconductivity_2016,wu2019ground}.
This structural transition corresponds to the $n_1$ or $n_2$ type lattice distortion.
$T_c$ increases to 4~K until 5~GPa under compression\cite{wu2019ground}.
The point-contact measurements also reported that $T_c \approx$ 2.1 K using a hard contact tip is higher than the measured $T_c \approx$ 1.13 K using a soft tip.
Assuming that the hard tip induces higher pressure than the soft tip, the experimental results are consistent with our result that $T_c$ is enhanced with increasing $n_1$ or $n_2$ lattice distortion [Fig.~\ref{fig5:phase_map_1}].
The experiments reported that the superconductivity is either conventional\cite{yu2017fully,wu2019ground} or unconventional\cite{xing_superconductivity_2016} depending on the physical situations. 
From our analysis, the superconducting phase of Au$_2$Pb is expected to be either a conventional fully gapped or unconventional topological mirror superconductor with a gapless SABS depending on physical parameters.

Similarly, in Cd$_3$As$_2$, the structural phase transition occurs near $2.6$ GPa, resulting in a monoclinic lattice $C_{2h}$.
Then, a superconductivity emerges at $T_c \approx 1.8$ K under pressure higher than $8.5$ GPa.
This structural transition corresponds to $n_3$ or $n_4$ type lattice distortion.
When the pressure increases further, $T_c$ keeps increasing from $1.8$ K ($8.5$ GPa) to $4.0$K ($21.3$ GPa), which is consistent with the enhancement of $T_c$ under lattice distortions [Fig.~\ref{fig6:phase_map_2}].
In this case, $n_1$ or $n_2$ can also be added without breaking the 
symmetry further.
From our analysis, the superconducting phases of Cd$_3$As$_2$ are expected to be either a conventional or topological mirror superconductor with a gapless SABS.
We emphasize that the topological nodal superconductor having a flat SABS can appear only if either $n_3$ or $n_4$ lattice distortion is turned on.
The point-contact measurements for Cd$_3$As$_2$ showed the zero-bias conductance peak (ZBCP) and double conductance peaks symmetric around zero bias, which was interpreted as a signal of a Majorana surface states\cite{aggarwal_unconventional_2016,he_pressure-induced_2016,wang2016observation}.
Even though our result cannot directly explain the result of the point-contact measurement, the unconventional superconductivity having gapless Majorana fermion can emerge regardless of the lattice distortions according to the surface spectra [see Fig.~\ref{fig4:surface_spectrum}], which seems to support the measured conductance peaks.
Further experimental studies that reveal the nature of superconductivity are necessary, and our theoretical results will be a helpful guideline to interpret the experimental result and search for the possible topological superconductivity in DSM.

\section*{Summary}
In this work, we have studied the possible symmetry-lowering lattice distortions and their effects on the emergence of unconventional superconductivity in doped topological DSM.
From the group theoretical analysis, four types of symmetry-lowering lattice distortions that reproduce the crystal systems
present in experiments are identified.
We investigated the possible superconductivity under such symmetry-lowering lattice distortions considering inter-orbital and intra-orbital electron density-density interactions.
We found that both conventional and unconventional superconductivity can emerge depending on the lattice distortion and electron density-density interaction.
Remarkably, the unconventional inversion-odd-parity superconductivity hosts gapless surface Andreev bound states (SABS) even under lattice distortions.
We found that the lattice distortion enhances the superconducting critical temperature.
Therefore, our work is consistent with the observed 
structural phase transition and the enhancement of superconductivity in Cd$_3$As$_2$ and Au$_2$Pb under pressure.
We also suggest that enhanced conventional and unconventional superconductivity in doped topological DSM can be controlled by physical parameters such as the pressure and strength of the superconducting pairing interaction.
Thus, our work will provide a valuable tool to explore and control the superconductivity in topological materials.

\section*{Methods}
To study the effects of symmetry-lowering lattice distortions, we assume a minimal $4 \times 4$ Hamiltonian that describes representative topological Dirac semimetals\cite{wang2013three,yang2014classification}, where the lattice distortions are implemented as a perturbation\cite{stoneham2001theory}.
To study the superconductivity, we construct the Bogoliubov-de Gennes (BdG) Hamiltonian within the mean-field approximation while keeping TRS and the crystal symmetry\cite{alexandrov2003theory,bennemann2008superconductivity}. 
The momentum independent pairing potentials are classified using irreducible representations of the unbroken point group \cite{bennemann2008superconductivity,fu2010odd,nakosai2012topological,kobayashi2015topological,hashimoto2016superconductivity}.
The nodal structures, chiral winding number in Eq.~(\ref{eq:chiral_winding_number_def}), and chiral mirror winding number in Eq.~(\ref{eq:mirror_chiral_winding_number}) are calculated using the BdG Hamiltonian.
The surface Green's functions are calculated using a Möbius transformation-based method\cite{umerski1997closed}.
The superconducting critical temperature $T_c$ is calculated by solving the linearized gap equation in the weak-coupling limit\cite{alexandrov2003theory,bennemann2008superconductivity,fu2010odd,nakosai2012topological,hashimoto2016superconductivity}.
All the details are provided in the main text and Supplementary Information.

\bibliography{reference}

\section*{Acknowledgements}
S.C. was supported by National  Research  Foundation  (NRF)  of Korea  through  Basic Science  Research  Programs (No. 2018R1C1B6007607, No. 2021R1H1A1013517), the research fund of Hanyang University (HY-2017), and the POSCO Science Fellowship of POSCO TJ Park Foundation.
K.H.L. was supported by the National Research Foundation of Korea(NRF) grant funded by the Korea government (MSIT) (No. 2021R1C1C1008738).
S.B.C. was supported by the National Research Foundation of Korea (NRF) grants funded by the Korea government (MSIT) (No. 2020R1A2C1007554) and the Ministry of Education
(No. 2018R1A6A1A06024977).
B-J.Y. was supported by the Institute for Basic Science in Korea (Grant No. IBS-R009-D1), Samsung Science and Technology Foundation under Project Number SSTF-BA2002-06, the National Research Foundation of Korea (NRF) grant funded by the Korea government (MSIT) (No. 2021R1A2C4002773).

\section*{Author contributions statement}
B-J. Y. conceived the project.  S. C. and K. H. L. performed the theoretical and numerical calculations. All authors analyzed the results. S. C. and K. H. L. wrote the manuscript. All authors reviewed the manuscript. 

\section*{Competing Interests} 
The authors declare no competing financial interests.

\section*{Additional information}


\noindent
\textbf{Correspondence} and requests for materials should be addressed to S.B.Chung or B-J. Yang.


\includepdf[pages=-]{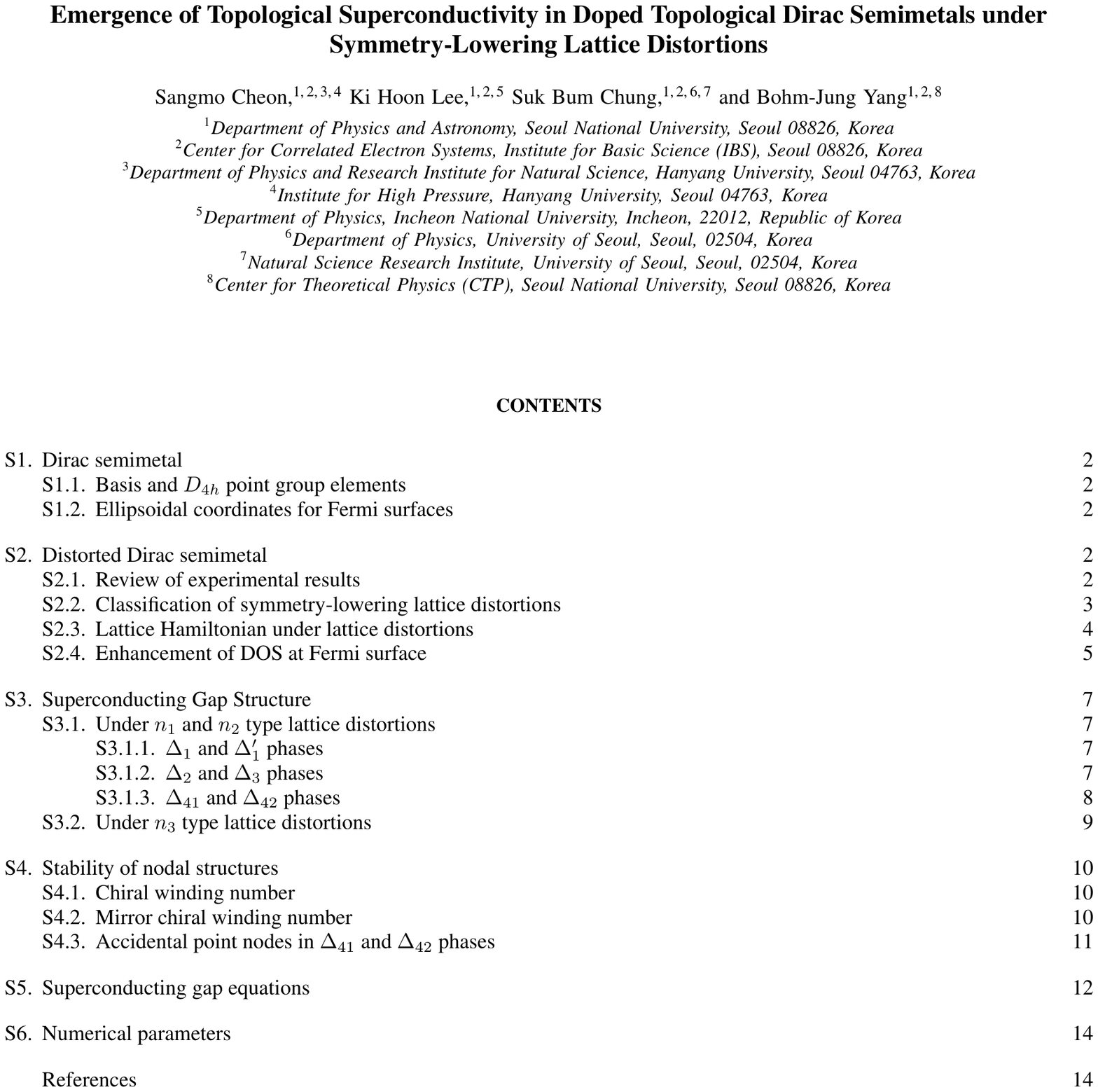}

\end{document}